\documentclass[12pt]{article}
\RequirePackage[OT1]{fontenc}
\RequirePackage{amsthm,amsmath}
\RequirePackage{natbib}
\RequirePackage[colorlinks,citecolor=blue,urlcolor=blue]{hyperref}

\usepackage{subfigure}
\usepackage{amssymb}
\usepackage{graphicx}
\usepackage{graphicx}
\usepackage{comment}
\usepackage{mathrsfs}
\usepackage[colorinlistoftodos]{todonotes}
\usepackage{rotating}
\usepackage{lscape}
\usepackage{epstopdf}
\usepackage{natbib}
\newtheorem{theorem}{Theorem}
\newtheorem{corollary}{Corollary}

\numberwithin{equation}{section}
\theoremstyle{plain}

\usepackage{enumerate}   
\usepackage{url} % not crucial - just used below for the URL 

\newcommand{\blind}{1}

\begin{document}

\def\spacingset#1{\renewcommand{\baselinestretch}%
{#1}\small\normalsize} \spacingset{1}

\if1\blind
{
  \title{\bf Bayesian time-aligned factor analysis of paired multivariate time series}
  \author{Arkaprava Roy$^1$ %1\thanks{
   % The authors gratefully acknowledge \textit{please remember to list all relevant funding sources in the unblinded version}}\hspace{.2cm}\\
    %Department of YYY, University of XXX\\
    and
    Jana Schaich-Borg$^2$
    and 
    David B Dunson$^1$ \\
    $^1$Department of Statistical Science,$^2$Social Science Research Institute, \\Duke University}
  \maketitle
} \fi

\if0\blind
{
  \bigskip
  \bigskip
  \bigskip
  \begin{center}
    {\LARGE\bf Bayesian time-aligned factor analysis of paired multivariate time series}
\end{center}
  \medskip
} \fi

\bigskip
	
	\begin{abstract}
		Many modern data sets require inference methods that can estimate the shared and individual-specific components of variability in collections of matrices that change over time. Promising methods have been developed to analyze these types of data in static cases, but very few approaches are available for dynamic settings. To address this gap, we consider novel models and inference methods for pairs of matrices in which the columns correspond to multivariate observations at different time points. In order to characterize common and individual features, we propose a Bayesian dynamic factor modeling framework called Time Aligned Common and Individual Factor Analysis (TACIFA) that includes uncertainty in time alignment through an unknown warping function.  We provide theoretical support for the proposed model, showing identifiability and posterior concentration.  The structure enables efficient computation through a Hamiltonian Monte Carlo (HMC) algorithm.  We show excellent performance in simulations, and illustrate the method through application to a social synchrony experiment.
	\end{abstract}
	
	\noindent%
	
	\vfill
\newpage
\spacingset{1.45} % DON'T change the spacing!
	
	\section{Introduction}
	Matrix-variate data are routinely collected in many fields.  As the collection of these types of data expands, so does the need for new statistical methods that can capture the shared and individual-specific structure in multiple matrices.  This is particularly the case when matrices in a collection consist of multivariate observations collected over time. Here, we are motivated by the challenge of measuring social coordination between two people who are interacting with one another.  In such cases, multiple facial and body features are extracted from videos of each individual in the pair over time.  The data for each individual form a matrix, with the columns corresponding to different time points. One component of the variability in the two matrices will be attributable to shared structure that relates to how much people like each other and cooperate \citep{lakin2003using,johnston2002behavioral,marsh2016imitation}.  Another component will be attributable to variability specific to each individual.  Analyzing these paired dynamic matrix-variate data requires a strategy that can accommodate two significant challenges: 1) complex multivariate dependence among variables, and 2) dynamic time-varying lags between the two multivariate time series.  Although our motivating example is from human social interactions, similar challenges are posed by other types of paired multivariate data, such as that collected in animal behavior studies, cellular imaging studies, finance, or speech, gesture, and handwriting recognition.
	
	%Time series data on interactions between individuals can be routinely collected using current video processing technology. This is not just limited to human interactions but applies to behavioral studies of animals and even basic  sequence studies of interactions between pairs of cells or chemicals and also in experiments such as speech recognition, handwriting recognition, gesture recognition etc. However, to the best of our knowledge, there is essentially no relevant statistical literature. Key challenges in analyzing these paired dynamic matrix-variate data are: 1) the complex multivariate dependence among the variables, and 2) the time-varying lag across the two multivariate time series. 
	
Joint and Individual Variation Explained (JIVE) \citep{lock2013joint} and Common and Individual Feature Analysis (CIFA) \citep{zhou2016group} were developed to capture shared and individual-specific features in pairs of multivariate matrices.  In the case of JIVE, the data $X_i$'s are decomposed into three parts: a low-rank approximation of joint structure $J_i$, a low-rank approximation of individual variation $S_i$, and an error $E_i$ under the restriction $JS_i^T=0$ for all $i$. Here $J$ is the matrix stacking $J_i$'s on top of each other. The CIFA decomposition defines a matrix factorization problem: $\min_{A, A_i,B_i,\tilde{B}_i}\|Y_i-(A, A_i)^T(B_i, \tilde{B}_i)\|_F^2$ under the restriction that $A^TA_i=0$ for all $i$, with $\|\cdot\|_F$ denoting the Frobenius norm. Thus JIVE assumes orthogonal rows between $J$ and $S_i$ whereas CIFA assumes orthogonality between columns of $A$ and $A_i$. Extensions of these methods are proposed in \cite{li2018general} and \cite{feng2018angle}. Related approaches have been used in behavioral research \citep{schouteden2014performing},
genomic clustering \citep{lock2013bayesian,ray2014bayesian}, neuroimaging \citep{zhou2016linked} and railway network analysis \citep{jere2014extracting}. In most cases, frequentist frameworks are used for inference, the methods are not likelihood-based, and the focus is on static data. \cite{de2018bayesian} developed a method for multigroup factor analysis in a Bayesian framework, which has some commonalities with these approaches but does not impose orthogonality.
	
	One way to accommodate time-varying lags is to temporally align the features in a shared space, avoiding the need to develop a complex model of lagged dependence across the series. However, time alignment is a hard problem. Typically, alignment is done in a first stage, and then an inferential model is applied to the aligned data \citep{vial2009combination}. However, such two-stage approaches do not provide adequate uncertainty quantification.
	
	%We consider to align the two time series by mapping the features of the shared space using a monotone increasing function $M:[0,1]\rightarrow [0,1]$, called a warping function. One interesting aspect of our motivating application is that the direction of mimicry might change over time. Sometimes the first person might be leading and then the second person takes charge. Since we model our warping function as a global monotone increasing function, it is flexible enough to capture such variations.
	
	Several approaches have been proposed to model warping functions. \cite{tsai2013profile} used basis functions similar to B-splines with varying knot positions, using stochastic search variable selection for the knots. This makes the model more flexible but at the cost of very high computational demand. \cite{kurtek2017geometric} put a prior on the warping function based on a geometric condition and developed importance sampling methods. In a multivariate setup, the method becomes very complicated due to the geometric structure. \cite{lu2017bayesian} use a similar structure in placing a prior on the warping function. However, the issue with computational complexity in the multivariate setting persists. 
	
	\cite{bharath2017partition,cheng2016bayesian} put
	a Dirichlet prior on the increments of the warping function over a grid of time points. Thus, the estimated warping function is not smooth. Also when the warping function is convolved with an unknown function, computation becomes inefficient due to poor mixing. The concept of warplets of \cite{claeskens2010multiresolution} is very interesting. Nevertheless, this method also suffers from a similar computational problem. 
	
	For multivariate time warping, \cite{listgarten2005multiple} proposed a method based on a hidden Markov  model. Other works propose to use a warping based distance to cluster similar time series \citep{orsenigo2010combining,che2017decade}. Unfortunately, these algorithms require the two time series to be collected at the same time points.  In addition, it is difficult to avoid a two-stage procedure, since there is no straightforward way to combine a statistical model with the warping algorithms. 
	
	\cite{gervini2004self} modeled the warping function as $M(t)=t+\sum_{j}s_jf_j(t)$, where $f_j(t)$'s are characterized using B-splines with the sum of the $s_j$'s equal to zero. For identifiability, they assumed restrictive conditions on the spline coefficients and did not accommodate multivariate data. \cite{telesca2008bayesian} developed a related Bayesian approach but their structure makes it difficult to apply gradient-based Metropolis-Hastings (MH), and finding a good proposal for efficient sampling is problematic. %for a model similar to \cite{gervini2004self}. Within their scheme of modeling, it is not straightforward to develop gradient-based Metropolis-Hastings (MH) algorithms for the parameters having no conjugacy structure. They proposed a classical MH-based algorithm, but finding a good proposal density for efficient sampling is problematic.% Apart from proposing an efficient computational scheme, our identifiability result is more general. Thus our proposed modeling of the warping function is much less restrictive. %Our novel modeling of the warping function allows us to use gradient-based MH sampling that ensures better mixing and estimation. Also in our model, for identifiability we do not require any monotonicity restriction on the latent factors. This makes our result stronger.
	
We propose to estimate the similarity between two multivariate time series with time-varying lags using a Bayesian dynamic factor model that incorporates time warping and parameter estimation in a single step. We assume the multivariate time series have both shared time-aligned factors and individual-specific dynamic factors. The resulting model reduces to a CIFA-style dependence structure, but unlike previous work, we accommodate time dependence and take a Bayesian approach to inference. Key aspects of our Bayesian implementation include likelihood-based estimation of shared and individual-specific subspaces, incorporation of a monotonicity constraint on the warping function for identifiability, and development of an efficient gradient-based Markov chain Monte Carlo (MCMC) algorithm for posterior sampling. 

We align the two time series by mapping the features of the shared space using a monotone increasing warping function $M : [0, 1] \rightarrow [0, 1]$. This flexible function can accommodate situations where the time lags between the multivariate time series change sign and direction.  Our monotone function construction differs from previous Bayesian approaches \citep{ramsay1988monotone,he1998monotone,neelon2004bayesian,shively2009bayesian,lin2014bayesian}, motivated by tractability in obtaining a nonparametric specification amenable to Hamiltonian Monte Carlo (HMC) sampling. %some shared set of dynamic state vector using a warping function  to capture shared or common features for dynamic matrix-variate data along with few more individual-specific dynamic latent factors in a Bayesian framework. To the best of our knowledge, this is the first Bayesian model for CIFA or JIVE class. We also develop efficient gradient-based MH sampling for this model. Apart from that, the warping function is convoluted with the shared set of time varying latent factors. Due to unavailability of conjugacy structure for the parameters in warping function in our modeling framework, we intend to use a gradient-based MH sampling for better mixing. Although there exists a vast literature on modeling smooth monotone increasing functions, it is not easy to calculate the derivative of the function for all of those constructions. In Bayesian framework, there are several works based on some modified splines, Gaussian process and some even proposed new basis functions to model monotone increasing functions \citep{ramsay1988monotone,he1998monotone,neelon2004bayesian,shively2009bayesian,lin2014bayesian}. Unfortunately, it is not straightforward to calculate the derivative of the function for most of these constructions. There is another approach to construct a monotone increasing function using the B-splines basis with coefficients, increasing in index \citep{de2001practical}. Several authors have used this approach for the modeling of monotone increasing functions \citep{brezger2008monotonic,telesca2008bayesian,roy2017high}. We also model the warping function using B-spline series prior with increasing coefficients, bounded between $[0, 1]$. We use the average cumulative sum of exponentiated real-valued numbers $\in (-\infty,\infty)$ to model the increasing coefficients. Our this novel nonparametric formulation of the warping function enables us to use Hamiltonian Monte Carlo (HMC) sampling \citep{girolami2011riemann}. This ensures better mixing. We also develop an efficient Bayesian computation for JIVE using HMC sampler. 
	
	In general, posterior samples of the loading matrices are not interpretable without identifiability restrictions \citep{seber2009multivariate,lopes2004bayesian,rovckova2016fast,fruehwirth2018sparse}. To avoid arbitrary constraints, which complicate computation, one technique is to post-process an unconstrained MCMC chain. \cite{assmann2016bayesian} post-process by solving an Orthogonal Procrustes problem to produce a point estimate of the loading matrix, but without uncertainty quantification. We consider to post-process the MCMC chain iteratively so that it becomes possible to draw inference based on the whole chain. Apart from the computational advantages, we also show identifiability of the warping function in our factor modeling setup  both in theory and simulations. Moreover, our identifiability result is more general than the result in \cite{gervini2004self} as we do not assume any particular form of the warping function other than monotonicity. %Our model can even be generalized for non-linear state-space models as well as for multiple sets of signals. 
	
	In section~\ref{model} we discuss our model in detail. Prior specifications are described in Section~\ref{prior}. Our computational scheme is outlined in Section~\ref{compute}. Section~\ref{theory} discusses theoretical properties such as identifiability of the warping function and posterior concentration. We study the performance of our method in two simulation setups in Section~\ref{sim}. Section~\ref{real} considers applications to human social interaction datasets. We end with some concluding remarks  in Section~\ref{discuss}. Supplementary Materials have all the proofs, additional algorithmic details, and additional results.
	
	\section{Modeling}
	\label{model}
	
	We have a pair of $p$ dimensional time varying random variables $X_t$ and $Y_t$. We propose to model the data as a function of time varying shared latent factors, $\eta(t)=\{\eta_1{(t)},\ldots,\eta_r{(t)}\}$, and individual-specific factors, $\zeta_1(t)=\{\zeta_{11}(t),\ldots,\zeta_{1r_1}(t)\}$ and $\zeta_2(t)=\{\zeta_{21}(t),\ldots,\zeta_{2r_2}(t)\}$. We do time alignment through the shared factors in $\eta(t)$ using warping functions $M_1(t),\ldots,$ $M_r(t)$. Here $M_i$ is the warping function for the latent variable $\eta_i$.  
	
	Latent factor modeling is natural in this setting in relating the measured multivariate time series to lower-dimensional characteristics, while reducing the number of parameters needed to represent the covariance. Since we are using the warping function to align the dynamic factors of the shared space, to ensure identifiability, the individual-specific space and the shared space space are required to be orthogonal. Thus the corresponding loading matrices of the two orthogonal subspaces are assumed to have orthogonal column spaces. Let $\Lambda$ be the loading matrix of the shared space. Then the shared space signal belongs to the span of the columns of $\Lambda$ with weights as some multiple of the shared factors $\eta(t)=\{\eta_1{(t)},\ldots,\eta_r{(t)}\}$. An element from the dynamic shared space can be represented as $\sum_{j=1}^{r}a_j\Lambda_{.j}\eta_{j}(t)$ for some constant $(a_1,\ldots, a_r)\in \mathbb{R}^r$ where $\Lambda_{.j}$ is the $j$-$th$ column of $\Lambda$. Alternatively it can also be written as $\Lambda \Xi_1 \eta(t)$, where $\Xi_1$ is a diagonal matrix with entries $(a_1,\ldots, a_r)$. The individual-specific space is assumed to be in the orthogonal subspace of the column space of $\Lambda$. Thus we use the orthogonal projection matrix $\Psi=I-\Lambda(\Lambda^{T}\Lambda)^{-1}\Lambda^T$ to construct the loading matrix of the individual-specific part of each signal. The loading matrix for the individual-specific space $X_t$ is assumed to be $\Psi \Gamma_1$ for some matrix $\Gamma_1$ of dimension $p\times r_1$, where $r_1$ is the rank. The corresponding loading matrix for the individual-specific space of $Y_t$ is $\Psi \Gamma_2$, with $\Gamma_2$ being a $p\times r_2$ matrix with $r_2$ the rank. 
	\begin{comment}
	\begin{align}
	X_{t} =& h_1(\eta_1{(t)},\ldots,\eta_q{(t)}) + \epsilon_1,\label{genmodel}\\
	Y_{t} =& h_2(\eta_1{(M_1(t))},\ldots,\eta_q{(M_q(t))}) + \epsilon_2,\nonumber
	\end{align}
	where $M_i$ is the warping function for the latent variable $\eta_i$ and $\epsilon_1\sim\mathrm{N}(0, \sigma_1^2)$, $\epsilon_2\sim\mathrm{N}(0, \sigma_2^2)$. We can model $h_1(\cdot)$ and $h_2(\cdot)$ using B-splines. 
	\end{comment}
	
\begin{figure}[htbp]
		\centering
		\subfigure[The direction of mimicry changes]{\label{fig:a}\includegraphics[width=60mm]{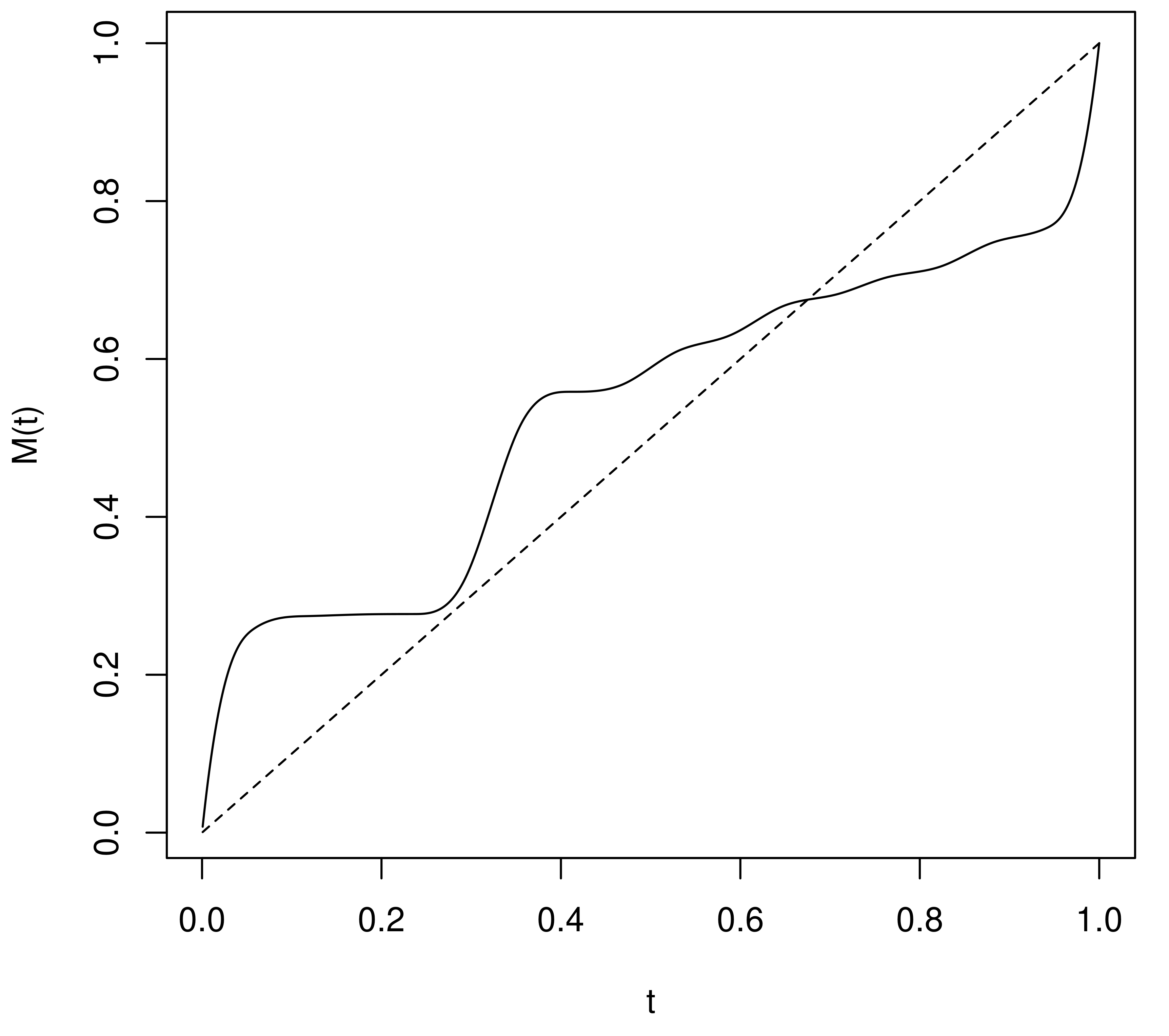}}
		\subfigure[The direction of mimicry does not change]{\label{fig:b}\includegraphics[width=60mm]{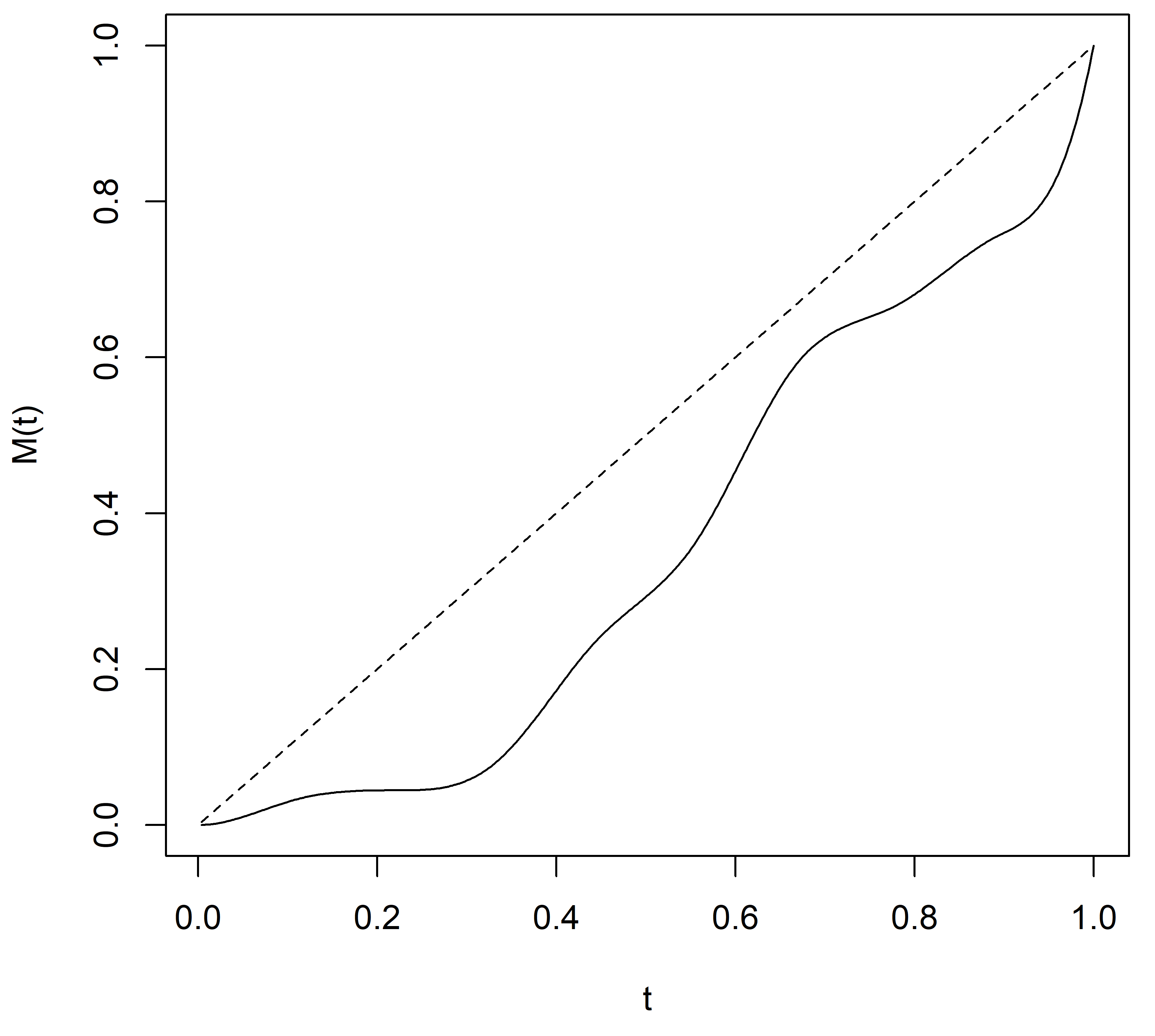}}
		\caption{Estimated warping functions for two social mimicry experiments (solid lines). The dashed lines correspond to the case in which no time alignment is needed.}
		\label{warpingex1}
	\end{figure}

	The warping function $M_i:[0,1]\rightarrow [0,1]$ is assumed to be monotone increasing, which is important for identifiability. To motivate, consider the case of social interactions.  People often imitate each other subconsciously.  In a normal conversation, people take turns mimicking each other without knowing it.  A method that models this mimicry would need to be able to account for the fact that the roles change dynamically over time. In Figure~\ref{warpingex1}, the dashed line through the origin with slope one corresponds to the case in which no alignment is needed. Panel (a) shows the warping function when one individual is mimicking the other for the first part of the experiment, and then the leader shifts. In the panel (b) experiment, the leader remains the same throughout. Both of these functions are estimated based on real data. 
	
	To model a smooth monotone increasing warping function bounded in [0, 1] such that $M_i(0)=0$ and $M_i(1)=1$ we use a B-spline expansion as follows,
	\begin{align*}
		M_i(t)=&\sum_{j=1}^J\gamma_{ij} B_j(t),
		\gamma_{ij}=\frac{\sum_{l=2}^j\exp(\kappa_{il})}{\sum_{k=2}^J \exp(\kappa_{ik})}, \quad \gamma_{i1}=0, 
	\end{align*}%\kappa_{ij}\sim \mathrm{N}(0, \phi),
	where $B_j(\cdot)$'s are B-spline basis functions and $\kappa_{ik}$ $\in (-\infty,\infty)$. To restrict $M_i(t)$ to be monotone increasing and bounded between [0, 1], it is sufficient to have the B-spline coefficients $\{\gamma_{ij}\}_{j=1}^J$ be monotone increasing in index $j$ and bounded between [0, 1] \citep{de2001practical}. This construction restricts $M_i$ to be a smooth monotone increasing function such that $M_i(0)=0$ and $M_i(1)=1$. These are the desired properties of a warping function. 
	
	For simplicity, we consider a single warping function for all the shared latent variables. 
	The complete model that we consider is
	\begin{align}
		X_t=&\Psi\Gamma_1\zeta_1(t)+\Lambda\Xi_1\eta({t})+\epsilon_{1t},\nonumber\\
		Y_{t}=&\Psi\Gamma_2\zeta_2(t)+\Lambda\Xi_2\eta(M(t))+\epsilon_{2t},\label{linearmodel}\\
		\zeta_{ij}(t)=&\sum_{j=1}^{K_i}\beta_{ilj} B_j(t),\quad i=1,2; l = 1,\ldots r_i\nonumber\\
		\eta_i(t)=&\sum_{j=1}^K\beta_{ij} B_j(t),\nonumber\\
		M(t)=&\sum_{j=1}^J\gamma_j B_j(t), \quad\gamma_j=\frac{\sum_{l=2}^j\exp(\kappa_l)}{\sum_{k=2}^J \exp(\kappa_k)},\quad \gamma_1=0,\nonumber\\
		\epsilon_i\sim & \mathrm{N}(0, \Sigma_i),\quad\Sigma_i=\mathrm{diagonal}(\sigma_{i1}^2,\ldots,\sigma_{ip}^2),\nonumber
	\end{align}
	where $\Lambda$, $\Gamma_1,\Gamma_2$ are static factor loading matrices of dimension $p\times r$,$p\times r_1$ and $p\times r_2$, respectively, with $\Psi=I-\Lambda(\Lambda^{T}\Lambda)^{-1}\Lambda^T$; $\Xi_1$ and $\Xi_2$ are $r\times r$ diagonal matrices; $r$ is the number of shared time varying latent factors and $r_1$, $r_2$ are the number of individual-specific latent factors for the 1st and 2nd individual, respectively; the error variances are given by $\Sigma_1$ and $\Sigma_2$. 
	
	We project the individual-specific loading matrices on the orthogonal space of the shared space spanned by columns of $\Lambda$ using $\Psi$. The data are collected over $T_1$ and $T_2$ time points longitudinally for individual 1 and 2 respectively, and $X$ and $Y$ are $p\times T_1$ and $p\times T_2$ dimensional data matrices. Correspondingly, $\Psi\Gamma_1\zeta$ and $\Lambda\Xi_1\eta$ are the individual-specific mean and shared space mean of $X$, respectively. The columns of these two matrices are orthogonal due to the orthogonality of $\Psi$ and $\Lambda$. Since $\zeta_1(t)$ and $\eta(t)$ are modeled independently, the rows of the two means are also independent in probability. A similar result holds for $Y$. Thus, this model conveniently explains both joint and individual variations.
	%We can assume $\Xi_1$ is the identity matrix due to inherent non-identifiability.
	
	The loading matrix $\Lambda$ identifies the shared space of the two signals. We assume a single shared set of latent factors $\eta(t)$ for both $X_t$ and $Y_t$. The warping function $M(t)$ aligns those for the $Y_t$ series relative to the $X_t$ series. Then we have individual-specific factors $\zeta_1(t),\zeta_2(t)$ and factor loading matrices $\Psi\Gamma_1,\Psi\Gamma_2$ that can accommodate within series covariances in $X(t)$ and $Y(t)$. We call our proposed method Time Aligned Common and Individual Factor Analysis (TACIFA).

	\section{Prior specification}
	\label{prior}
	We use priors similar to those in \cite{bhattacharya2011sparse} for $\Lambda$, $\Gamma_1$ and $\Gamma_2$ to allow for automatic selection of rank. We try to maintain conjugacy as much as possible for easier posterior sampling. The detailed prior description for $\kappa,\beta,\Lambda, \Xi_1,\Xi_2,\Gamma_1,\Gamma_2,\sigma_1$ and $\sigma_2$ is described below,
	$$\Lambda_{lk}|\phi_{1,lk},\tau_{1k}\sim \mathrm{N}(0,\phi_{1,lk}^{-1}\tau_{1k}^{-1}),$$
	$$\phi_{1,lk}\sim \mathrm{Gamma}(\nu_1,\nu_1),\quad \tau_{1k}=\prod_{i=1}^k\delta_{mi}$$
	$$\delta_{1,1}\sim \mathrm{Gamma}(\alpha_{1}, 1),\quad\delta_{1,i}\sim \mathrm{Gamma}(\alpha_{2}, 1),$$ for $1\leq i\leq r$,
	$$\Gamma_{1,lk}|\phi_{11,lk},\tau_{11k}\sim \mathrm{N}(0,\phi_{11,lk}^{-1}\tau_{11k}^{-1}),\quad \Gamma_{2,lk}|\phi_{12,lk},\tau_{12k}\sim \mathrm{N}(0,\phi_{12,lk}^{-1}\tau_{12k}^{-1}),$$ 
	$$\phi_{1n,lk}\sim \mathrm{Gamma}(\nu_1,\nu_1),\quad \tau_{1nk}=\prod_{i=1}^k\delta_{mi}$$
	$$\delta_{1n,1}\sim \mathrm{Gamma}(\alpha_{1n1}, 1),\quad\delta_{1n,i}\sim \mathrm{Gamma}(\alpha_{1n2}, 1),$$ for $1\leq i\leq r_1$ for $n=1$,$1\leq i\leq r_2$ for $n=2$,
	$$\sigma_{1l}^{-2}\sim\mathrm{Gamma}(\alpha_1,\alpha_1),\quad\sigma_{2l}^{-2}\sim\mathrm{Gamma}(\alpha_2,\alpha_2),$$ for $1\leq l\leq p$ and
	$$\Xi_{1,ll},\Xi_{2,ll},\kappa_j,\beta_{qk}\beta_{siK_s}\sim N(0, \omega).$$ for $1\leq k\leq K$,$q=1,\ldots,r$ $1\leq j\leq J$, $i=1,\ldots,r_s$,$s=1,2$ and $l=1,\ldots,r$.
	%$$\Pi[K=k]= b_1'\exp[-b_2 k (\log k)^{b_3}],\Pi[J=j]= b_1\exp[-b_2' j (\log j)^{b_3'}],$$
	%with $b_1,b_2, b_1',b_2'>0$ and $0\le b_3,b_3'\le 1$. Here for $b_3=0$ it would be a geometric distribution and for $b_3=1$, it would become a Poisson distribution. 
	Higher values of $\alpha_{m2}$ ensure increasing shrinkage as we increase rank.
	
	\section{Computation}
	\label{compute}
	We use Gibbs updates for all of the parameters except for $\Lambda$ and $\kappa$; details are provided in Section 2 of Supplementary Materials. For $\Lambda$ and $\kappa$, we propose an efficient gradient-based Metropolis-Hastings algorithm. For our proposed model, we can easily calculate the derivative of the log-likelihood with respect to $\kappa$ using derivatives of B-splines \citep{de2001practical}. This parameter $\kappa$ is only involved in the model of $Y_t$. The negative of that log-likelihood function including the prior on $\kappa$ is $$L(\kappa)=\sum_{t=1}^T\sum_{i=1}^p\frac{1}{\sigma_{2i}^2}\Bigg[Y_{it}-\Psi_{2i}\zeta_2(t)-\Lambda_{2i}\eta\Big\{\sum_{j=1}^J\frac{\sum_{l=2}^j\exp(\kappa_l)}{\sum_{k=2}^J \exp(\kappa_k)} B_j(t)\Big\}\Bigg]^2 + \frac{\sum_{j=2}^J\kappa_j^2}{2\phi^2}.$$
	For simplicity in expression of the derivative, let us denote $A_{it} \\= \Lambda_{2i}\eta\big(\sum_{j=1}^J\frac{\sum_{l=2}^j\exp(\kappa_l)}{\sum_{k=2}^J \exp(\kappa_k)} B_j(t)\big)$ and $M(t)=\sum_{j=1}^J\frac{\sum_{l=2}^j\exp(\kappa_l)}{\sum_{k=2}^J \exp(\kappa_k)} B_j(t)$, as defined earlier.
	Then the derivative is given by 
	\begin{align*}
		L'(\kappa_j)=&-\sum_{t=1}^T\sum_{i=1}^p\frac{1}{\sigma_{2i}^2}(Y_{it}-\Psi_{2i}\zeta_2(t)-A_{it})A_{it}\Bigg[\sum_{l=j}^{J}B_l(t)\\&\quad -M(t)\Bigg]\exp(\kappa_j)/\sum_{k=2}^J \exp(\kappa_k) + \kappa_j/\omega^2.
	\end{align*}
	Let us denote $L'(\kappa)=(L'(\kappa_2),\ldots,L'(\kappa_J))'.$ 
	
	Now, we discuss the sampling for $\Lambda$. To update the $j$-$th$ column of $\Lambda$, we first rewrite the orthogonal projection matrix using the matrix inverse result of block matrices as $$\Psi=(I-P_1)(I-P_2)(I-P_1)$$
	where $P_1=\Lambda_{.-j}(\Lambda_{-j}^T\Lambda_{-j})^{-1}\Lambda_{.-j}^{T}$ and $P_2=\Lambda_{.j}(\Lambda_{j}^T(I-P_1)\Lambda_{j})^{-1}\Lambda_{.j}^{T}$. Here $\Lambda_{.-j}$ is the reduced matrix after removing the $j$-$th$ column of $\Lambda$ and $\Lambda_{. j}$ is the $j$-$th$ column. The negative log-likelihood with respect to $\Lambda_{. j}$ is
	
	\begin{align*}
		L_1(\Lambda_{.j})&=\sum_{t}\sum_{i=1}^{p}(X-\Psi\Gamma_1\zeta_1(t)-\Lambda\Xi_1\eta(t))^2/(2\sigma_1^2) \\&\quad+\sum_{t}\sum_{i=1}^{p}(Y-\Psi\Gamma_2\zeta_1(t)-\lambda\Xi_2\eta(t))^2/(2\sigma_2^2)+\sum_{k}\Lambda_{kj}^2/(2\phi_{1,kj}\tau_j),
	\end{align*}
	and the derivative is 
	\begin{align*}
		L_1'(\Lambda_{kj})&=\sum_{t}\sum_{i=1}^{p}(X_{ti}-\Psi\Gamma_1\zeta_1(t)-\Lambda\Xi_1\eta(t))(B_{ti}-\eta(t))/(\sigma_1^2)+\sum_{t}\sum_{i=1}^{p}(Y_{ti}-\\&\quad\Psi\Gamma_2\zeta_2(t)-\Lambda\Xi_2\eta(M(t)))(B_{ti}-\eta(t))/(\sigma_2^2)+\Lambda_{kj}/(\phi_{1,kj}\tau_j),
	\end{align*}
	where 
	\begin{align*}
	B = &-(I-P_1)Q(I-P_1)\\
		Q=&\bigg\{\big((\Lambda_{.j}e_k^T+e_k\Lambda_{.j}^T)\Lambda_{.j}^T(I-P_1)\Lambda_{.j}\\&\quad-2e_k(I-P_1)\Lambda_{.j}^T\Lambda_{.j}(I-P_1)\Lambda_{.j}^T\big)/(\Lambda_{.j}^T(I-P_1)\Lambda_{.j})^2\bigg\},
	\end{align*}
	with $e_k$ a vector of length $p$ having 1 at the $k$-$th$ position and zero elsewhere.
	
Relying on the above gradient calculations we use HMC \citep{duane1987hybrid, neal2011mcmc}. %The HMC algorithm uses Hamiltonian dynamics to derive proposals for the MH sampler. Instead of random walk it uses a momentum variable which is characterized by potential and kinetic energy functions. For potential energy, negative log-likelihood is used and as kinetic energy sum of square is a standard choice for normal priors. These energy functions can be expressed in the form of differential equations. To discretize those equations, a modified Euler’s method is applied. After that it is further modified using the leapfrog method. The leapfrog method involves two components, the number of leapfrog steps and the step length. The number of leapfrog steps is the number of times the Euler's method is applied to update the parameter before passing it to the MH step. The step length is step size for each update using Euler's method. Thus the effective step size is the product of the number of leapfrog steps and the step length. \cite{neal2011mcmc} has described the whole algorithm succinctly in the form of an R code in the Figure 2 of that material. We use that same algorithm for our estimation.
	We keep the leapfrog step fixed at 30. We tune the step size parameter to maintain an acceptance rate within the range of 0.6 to 0.8. If the acceptance rate is less than 0.6, we reduce the step length and increase it if the acceptance rate is more than 0.8. We do this adjustment after every 100 iterations. We also incorporate removal of columns of $\Lambda$, $\Gamma_1$ and $\Gamma_2$ if the contributions are below a certain threshold as described in Section 3.2 of \cite{bhattacharya2011sparse}.
	
	\begin{comment}
	{\it Initialization:} In our simulations, proper initialization of the parameters is found to be advantageous to get better results. We take the following strategy to initialize the parameters of our model.
	\begin{enumerate}
		\item Initialize $\Gamma_1$ and $\Gamma_2$ as zero matrices.
		\item Generate $\Lambda$ from the prior and set $\Xi_1$ and $\Xi_2$ as identity matrices
		\item Given the value of $K$, from the model $X_t=\Lambda\eta(t)+\epsilon_1$, we get an MLE estimate for $\eta$ given $\Lambda$.
		
		\item Now use the Bayes procedure using the assigned priors on the subsection of the model $Y_t=\Lambda\Xi_2\eta(M(t))+\epsilon_2$ to get an estimate for $M(t)$ and $\Xi_2$ given the estimated functions $\eta(t)$ and $\Lambda$ from previous step. In this step, we collect only 200 MCMC samples and use last 100 samples to get the initial estimates. This approach works very well in the simulations.
	\end{enumerate}
	\end{comment}

	\subsection{Post-MCMC inference}
	\label{postMCMC}
	Here we discuss the strategy to infer the loading matrix $\Lambda_1=\Lambda\Xi_1$. The loading matrices are identifiable up to an orthogonal right rotation. This implies that $(\Lambda_1, \eta)$ and $(\Lambda_1R, R^T\eta)$ for some orthonormal matrix $R$ have equivalent likelihood.
	
	Let $\Lambda_1^{(1)},\ldots,\Lambda_1^{(m)}$ be $m$ post burn-in samples of $\Lambda_1$. We post-process the chain successively moving from the first sample to the last. First $\Lambda_1^{(2)}$ is rotated with respect to $\Lambda_1^{(1)}$ using some orthonormal matrix $R_1$ such that $\|\Lambda_1^{(1)}-\Lambda_1^{(2)}R_1\|_F^2$ is minimized, where $\|\|_F^2$ denotes the Frobenius norm. This minimization criterion rotates $\Lambda_1^{(2)}$ to make it as close as possible to $\Lambda_1^{(1)}$. The solution of $R_1$ is obtained in Theorem~\ref{loadiden}. Then we post-process $\Lambda_1^{(3)}$ with respect to $\Lambda_1^{(2)}R_1$ and so on.
	
	\begin{theorem}
		The minimizer $R_1$ of the objective function $\|\Lambda_1^{(1)}-\Lambda_1^{(2)}R_1\|_F^2$ is given by $R_1=Q_2Q_1^{T}$, where $Q_1DQ_2^{T}$ is the singular value decomposition (SVD) of $(\Lambda_1^{(1)})^{T}\Lambda_1^{(2)}$. 
		\label{loadiden}
	\end{theorem}
	
	The proof of the theorem is in the Section 1.1 of Supplementary Materials. Intuitively, the columns of $Q_1$ and $Q_2$ are the canonical correlation components of $\Lambda_1^{(1)}$ and $\Lambda_1^{(2)}$, respectively. Thus the rotation matrix $R_1$ rotates $\Lambda_1^{(2)}$ towards the least principal angle between $\Lambda_1^{(2)}$ and $\Lambda_1^{(1)}$. For instance, $\Lambda_1^{(2)}$ could be an exact right rotation of $\Lambda_1^{(1)}$. Thus before starting to post-process the MCMC chain, we transform $\Lambda_1^{(1)}$ as $\Lambda_1^{(1)}U_2$ such that $U_1EU_2^T$ is the SVD of the residual $(X_t-\Psi^{(1)}\Gamma_1^{(1)}\zeta(t)^{(1)})^T\Lambda_1^{(1)}$ in the same way and here $E$ is the diagonal matrix with elements in decreasing order. This initial transformation ensures that the higher order columns of the loading matrix are lower in significance in explaining the data. Then following the above result, we post-process the rest of the MCMC chain of the loading matrix on the post burn-in samples successively. In general, SVD computation is expensive. However, in most applications, the estimated rank is very small. Thus the computation becomes manageable. After the post-processing, we can construct credible bands for the parameters.
	
	%We also develop a strategy to estimate the separation between individual-specific subspaces of $X$ and $Y$. The individual-specific spaces are characterized by $\Gamma_1$ and $\Gamma_2$. We calculate $$S_l=P(\mathrm{CC}(\Gamma_1, \Gamma_2)>l),$$ for some given $l$, with CC denoting the canonical correlation. This $l$ sets the threshold of similarity. For instance, if we have $l=\cos(\pi/4)$, then $S_l$ is the probability that the two subspaces are separated by principal angles smaller than $\pi/4$. For the $m$-$th$ sample we define $C_m=\mathrm{mean}\{I(\mathrm{CC}(\Gamma_1, \Gamma_2)>l)\}$, where $I$ is an indicator function. Thus we get a posterior estimate of $S_l$ as the average of $C_m$ over all post burn-in MCMC samples along with a measure of uncertainty.
	
	 \subsection{Measure of similarity}
    \label{simmeasure}
	It is of interest to quantify similarity between paired time series. % when both time series are defined with the same set of features. Shared space and warping are two separate components in our model. The shared space can have features with different warping functions. The size of the shared space between the two time series increases as more features in the shared space share similar warping functions.  The size of the shared space between the two time series decreases as the features and warping functions of the time series diverge.  Thus 
	We propose the following measure of similarity, 
	
	\begin{comment}
	$\text{Syn}(X, Y)$
	$$=1-\frac{1}{pT}\sum_{l=1}^p\Bigg|\sum_{t=1}^T\bigg[\frac{(\Lambda_l\Xi_1\eta({t}))^2}{(\Psi_l\Gamma_1\zeta_1(t))^2+(\Lambda_l\Xi_1\eta({t}))^2+\sigma_{1l}^2} - \frac{(\Lambda_l\Xi_2\eta(M({t})))^2}{(\Psi_l\Gamma_2\zeta_2(t))^2+(\Lambda_l\Xi_2\eta(M({t})))^2+\sigma_{2l}^2}\bigg]\Bigg|,$$ 
	\end{comment}
	
	$\text{Syn}(X, Y)$
	$$=1-\frac{1}{pT}\sum_{l}\Bigg|\sum_{t}\bigg[\frac{(\Lambda_l\Xi_1\eta({t}))^2}{(\Psi_l\Gamma_1\zeta_1(t))^2+(\Lambda_l\Xi_1\eta({t}))^2+\sigma_{1l}^2} - \frac{(\Lambda_l\Xi_2\eta(M({t})))^2}{(\Psi_l\Gamma_2\zeta_2(t))^2+(\Lambda_l\Xi_2\eta(M({t})))^2+\sigma_{2l}^2}\bigg]\Bigg|,$$ where $\Lambda_l$, $\Psi_l$ denote the $l^{th}$ row of the corresponding matrices and $p$,$T$ denote number of features and time points respectively. The measure `Syn' is bounded between $[0,1]$. Here, the difference in relative contribution of each feature on the two shared spaces is considered as a measure of disimilarity. Then as a measure of similarity, we consider the difference of the average disimilarity from one. Smaller Syn-value would suggest that the warping function is not able to align the shared space perfectly. %We calculate the similarity measure on those above mentioned three datasets. %with the corresponding values on a shuffled dataset that is expected to be warped poorly.

	%\cite{assmann2016bayesian} developed an iterative algorithm to get a posterior estimate from the unconstrained MCMC chain of the loading matrix by solving an orthogonal procrustes problem. We compare the posterior mean from our algorithm with this method in a small simulation study and show that performance of our algorithm is similar or at least better. The simulation is in the Supplementary materials (for now, will move to supplementary). The advantage of using our method is that we can recover the entire chain and it is a one pass algorithm with similar computational demand as in a single iteration of the orthogonal procrustes problem.
	
	\section{Theoretical support}
	\label{theory}
	In this section, we provide some theoretical justification for our model. Identifiability of the warping function is a desirable property as well as posterior consistency. 
	\subsection{Identifiability of the warping function}
	The following result shows that the warping function $M(t)$ is identifiable for model \eqref{linearmodel}.
	\begin{theorem}
		\label{identify}
		The warping function $M(t)$ is identifiable if $\eta(t)$ is continuous and not constant at any interval of time.
	\end{theorem}
	
	The proof is by contradiction. Details of the proof are in Section 1.2 of Supplementary Materials. The assumptions on $\eta(t)$ are very similar to those assumed for the `structural mean' in \cite{gervini2004self}. The continuity assumption of $\eta(t)$ can be replaced with a `piecewise monotone without
	flat parts' assumption \citep{gervini2004self}. The proof is still valid with minor modifications for this alternative assumption. In our model $\eta(t)$ is varying with time smoothly. Thus $M(t)$ is identifiable. %We would like to point out that this result can be generalized for the general model in~\eqref{genmodel} as well under some additional condition. If for $(h_1, h_2)$ of the general model there exists an invertible function $\Gamma$ such that $h_1(x)=h_2(\Gamma(x))$, it is possible to verify the above result. The techniques are similar to the proof of above Theorem~\ref{identi}.
	
	\subsection{Asymptotic result}
	We study the posterior consistency of our proposed model. Our original model is %$\Xi_{1},\Xi_2$ from the model in \eqref{model}. Thus we rewrite our model as 
	\begin{align}
		X_t=&\Psi\Gamma_1\zeta_1(t)+\Lambda\Xi_1\eta({t})+\epsilon_{1t},\quad \epsilon_{1t}\sim \mathrm{N}(0,\sigma_{1t}^2),\nonumber\\
		Y_{t}=&\Psi\Gamma_2\zeta_2(t)+\Lambda\Xi_2\eta(M(t))+\epsilon_{2t},\quad \epsilon_{2t}\sim \mathrm{N}(0,\sigma_{2t}^2).\label{thmmodel}
	\end{align}
	%At max $\Lambda\Xi_1$, $\Lambda\Xi_2$,$\Psi\Gamma_1$,$\Psi\Gamma_2$, can be of $p\times p$ dimensional matrices and in our asymptotic regime $p$ is kept fixed. Thus the posterior contraction rates will not be affected much due to these parameters when we have the infinite dimensional non-parametric functions $\zeta_1$, $\zeta_2$, $\eta$ and $M$. In this section, we drop these four parameters for simplified calculations. 
	We first show posterior concentration of a simplified model that drops $\Xi_1$ and $\Xi_2$. Then using that result we show posterior concentration of model \eqref{thmmodel} in Corollary~\ref{cor}. We rewrite $\zeta_1(t)=\Psi\Gamma_1\zeta_1(t)$, $\zeta_2(t)=\Psi\Gamma_2\zeta_2(t)$ and $\eta(t)=\Lambda\eta(t)$. Based on the constructions, $\zeta_i(t)$ and $\eta(t)$ are orthogonal for $i=1,2$. 
	We consider the following simplified model,
	\begin{align*}
		X_{t_i}=&\zeta_1(t_i)+\eta({t_i})+\epsilon_{1t_i},\quad \epsilon_{1t}\sim \mathrm{N}(0,\sigma_{1t_i}^2),\\
		Y_{t_i}=&\zeta_2(t_i)+\eta(M(t_i))+\epsilon_{2t_i},\quad \epsilon_{2t}\sim \mathrm{N}(0,\sigma_{2t_i}^2),
	\end{align*}
	for $0\leq t_i\leq 1$ and $i=1,\ldots,n$. We study asymptotic properties in the increasing $n$ and fixed $p$ regime. We need to truncate the B-spline series after a certain level or place a shrinkage prior on the number of B-splines as $\Pi[K=k]= b_1'\exp[-b_2' k (\log k)^{b_3'}],\Pi[J=j]= b_1\exp[-b_2 j (\log j)^{b_3}],$ $\Pi[K_i=j]= b_{i1}\exp[-b_{i2} j (\log j)^{b_{i3}}]$ for $i=1,2$,
	with $b_1,b_2,b_{12},b_{22} b_1',$ $b_2',b_{11},b_{21}>0$ and $0\le b_3,b_3',b_{13},b_{23}\le 1$. For $b_3=0$ we obtain a geometric distribution and for $b_3=1$, a Poisson distribution. 
	
	To study posterior contraction rates, we consider the empirical $\ell_2$-distance on the regression functions. The empirical $\ell_2$-distance for the two sets of parameters $(\zeta_{11},\zeta_{21},\eta_1,M_1)$ and $(\zeta_{12},\zeta_{22},\eta_2, M_2)$ is given by
	\begin{align*}
		&d^2((\zeta_{11},\zeta_{21},\eta_1,M_1), (\zeta_{12},\zeta_{22},\eta_2,M_2))\\
		&\quad = \frac{1}{n}\sum_{i=1}^n\big[\|\zeta_{11}(t_i)-\zeta_{12}(t_i)\|_2^2+\|\zeta_{21}(t_i)-\zeta_{22}(t_i)\|_2^2+\|\eta_1(t_i)-\eta_2(t_i)\|_2^2\\&\qquad+\|\eta_1(M_1(t_i))-\eta_2(M_2(t_i))\|_2^2\big].
	\end{align*}

	The smoothness of the underlying true functions $\zeta_{10},\zeta_{20},\eta_0$ and $M_0$ plays the most significant role in determining the contraction rate. The fixed dimensional parameters $\sigma_1$ and $\sigma_2$ do not have much impact on the rate. The constants $b_{13},b_{23},b_3$ and $b_3'$ appearing in the prior for the number of B-spline coefficients $K_1,K_2,K, J$ have a mild effect. 
	
	\begin{theorem}
		\label{concent}
		Assume that the true functions $\zeta_{10},\zeta_{20},\eta_0$ and $M_0$ belong to H\"older classes of smooth functions and are of regularity levels $\iota_1,\iota_2,\iota$ and $\iota'$ on $[0, 1]$. Then the posterior contraction rate is given by
		$$
		n^{-\bar{\iota}/(2\bar{\iota}+1)} (\log n)^{\bar{\iota}/(2\bar{\iota}+1)+(1-\bar{b_3})/2},
		$$
		where $\bar{\iota}=\min\{\iota,\iota_1,\iota_2,\iota'\}$ and $\bar{b}_3=\min\{b_3,b_3',b_{13},b_{23}\}$.
		%$$\max\bigg\{n^{-\iota/(2\iota+1)} (\log n)^{\iota/(2\iota+1)+(1-b_3)/2},n^{-\iota'/(2\iota'+1)} (\log n)^{\iota'/(2\iota'+1)+(1-b_3')/2}\bigg\}$$
	\end{theorem}
	The proof is based on the general theory of posterior contraction as in \cite{Ghosal} for non-identically distributed independent observations and results for finite random series priors \citep{shen2015adaptive}. Details of the proof are in Section 1.3 of Supplementary Materials.
	
	Let the parameter space for dynamic latent factors $\zeta_1,\zeta_2,\eta$ be $\mathcal{F}$, which is the class of real-valued smooth continuous functions on [0,1], and for the warping function $\mathcal{M}$ be the class of [0, 1] bounded smooth monotone continuous functions on [0,1]. Let $\tilde{X},\tilde{X},\tilde{L}, \tilde{G}_1, \tilde{G}_2$ be the priors for the matrices $\Xi_1, \Xi_2, \Lambda, \Gamma_1, \Gamma_2$, respectively, and $\mathcal{X},\mathcal{L}, \mathcal{G}_1, \mathcal{G}_2$ are the parameter spaces of $\tilde{X},\tilde{L}, \tilde{G}_1, \tilde{G}_2$, respectively.
	
	{\it Assumption 1}: For the true loading matrices and functions, we have 
	
	$\{\Xi_{10},\Xi_{20},\Lambda_0, \Gamma_{10}, \Gamma_{20},\zeta_{10},\zeta_{20},\eta_0,M_0\}\in \mathcal{X}^2\times\mathcal{L}\times\mathcal{G}_1\times\mathcal{G}_2\times\mathcal{F}^3\times\mathcal{M}.$ \\
	
	Similarly we can define empirical $\ell_2$-distance $d_1^2((\Psi_1, \Lambda_1, \Gamma_{11},\Gamma_{12},\Xi_{11},\Xi_{12}, \zeta_{11},\zeta_{21},\eta_1,M_1),$ $ (\Psi_2, \Lambda_2, \Gamma_{21},\Gamma_{22},\Xi_{21},\Xi_{22}, \zeta_{12},\zeta_{22},\eta_2,M_2))$ as $d^2$ for the full model and we have following consistency result.
	
	\begin{comment}
	\begin{align*}
		&d_1^2((\Psi_1, \Lambda_1, \Gamma_{11},\Gamma_{12},\Xi_{11},\Xi_{12}, \zeta_{11},\zeta_{21},\eta_1,M_1), (\Psi_2, \Lambda_2, \Gamma_{21},\Gamma_{22},\Xi_{21},\Xi_{22}, \zeta_{12},\zeta_{22},\eta_2,M_2))\\
		&\quad = \frac{1}{n}\sum_{i=1}^n\big[\|\Psi_1\Gamma_{11}\zeta_{11}(t_i)-\Psi_2\Gamma_{12}\zeta_{12}(t_i)\|_2^2+\|\Psi_1\Gamma_{21}\zeta_{21}(t_i)-\Psi_2\Gamma_{22}\zeta_{22}(t_i)\|_2^2\\&\qquad+\|\Lambda_1\Xi_{11}\eta_1(t_i)-\Lambda_2\Xi_{12}\eta_2(t_i)\|_2^2+\|\Lambda_1\Xi_{21}\eta_1(M_1(t_i))-\Lambda_2\Xi_{22}\eta_2(M_2(t_i))\|_2^2\big].
	\end{align*}
	\end{comment}
	
	\begin{corollary}
		\label{cor}
		Under the above assumption, the posterior for parameters in the model \eqref{thmmodel} is consistent with respect to the distance $d_1$.
	\end{corollary}

    For the full model in \eqref{thmmodel}, the test constructions will remain the same as in the proof of Theorem~\ref{concent}. We only need to verify the Kullback-Leibler prior positivity condition. Within our modeling framework, Assumption 1 trivially holds. Details of the proof are in Section 1.4 of Supplementary Materials. The posterior contraction rate of this full model will be the same as the given rate of Theorem~\ref{concent} as the loading matrices can at most be $p\times p$-dimensional and we assume $p$ is fixed.
	
	\section{Simulation Study}
	\label{sim}
	We run two simulations to evaluate the performance of TACIFA on pairs of multivariate time series. We evaluate TACIFA by: (1) ability to retrieve the appropriate number of shared and individual factors, (2) accuracy of the estimated warping functions and accompanying uncertainty quantification, (3) out of sample prediction errors, and (4) performance relative to two-stage approaches for estimating shared and individual-specific dynamic factors. In the first simulation, we generate data from the proposed model. In the second simulation, we analyze two shapes changing over time, data that does not have any inherent connection to our proposed model.
	
	To assess out of sample prediction error, we randomly assign 90\% of the data points to the training set and the remaining 10\% to the test set. The two-stage approaches we compare our method to apply JIVE on the training set in the first stage to estimate the shared space and warp the shared matrices, and then apply multivariate imputation algorithms ({\tt missForest, MICE, mtsdi}) in the second stage to make predictions on the testing data set. We evaluate the performance of naive dynamic time warping (based solely on minimization of Euclidean distance), derivative dynamic time warping (based on local derivatives of the time data to avoid singularity points), and sliding window based dynamic time warping. Since our model is the only approach with a mechanism for uncertainty quantification, we can compare the prediction performance of TACIFA to two-stage approaches, but we cannot compare uncertainty estimation.
	
	The individual-specific loading matrices are $\Psi\Gamma_1$ and $\Psi\Gamma_2$. The shared space loading matrices are $\Lambda\Xi_1$ and $\Lambda\Xi_2$. For the ($i,j$)-$th$ coordinate of a loading matrix $A$, we define a summary measure $SP_{i,j}(A)=\big(|0.5-P(A[i,j]>0)|\big)/0.5$ quantifying the ``importance" of the element. Here $P(A[i,j]>0)$ is the posterior probability estimated from the MCMC samples of $A$ after performing the post-processing steps defined in Section~\ref{postMCMC}. These scores help to quantify the importance of the factors and to estimate the number of important factors retrieved by the model.
	
	\subsection{Simulation case 1}
	First we generate the data from a finite factor model with the following specifications, $\zeta_{1k}(t)=\sin(kt)$, $\zeta_{2k}(t)=\cos(kt)$, $\eta_k(t) = \cos(kt)$ and $M_0(t)=t^{0.5}$, with $k$ varying from 1 to 10. The factor loading matrices are of dimension $20\times 10$, with the elements of $\Gamma_1$,$\Gamma_2, \Lambda$ generated independently from N$(0, 1)$. We have $p=20$, $r=10$. We vary $t$ from $1/500$ to 1 with an increment of $1/500$. The data $X_t$ and $Y_t$ are generated from N$(\Psi\zeta_1+\Lambda\eta(t), 1)$ and N$(\Psi\zeta_2+\Lambda\eta(M(t)), 1)$, respectively, where $\eta(t)=(\eta_1(t),\dots,\eta_{10}(t))$ and $\Psi=I-\Lambda(\Lambda^{T}\Lambda)^{-1}\Lambda^T$. 
	
	The choices of hyper parameters are $\phi=100$, $\alpha_{i1}=\alpha_{i2}=5$ for $i=1,2$, $K_1=K_2=K=10$ and $J=20$. The hyperparameters of the inverse gamma priors for the variance components are all $0.1$ which is weakly informative. We collect 6000 MCMC samples and consider the last 3000 as post burn-in samples for inferences.
	
\begin{figure}[htbp]
		\centering
		\includegraphics[width = 0.8\textwidth]{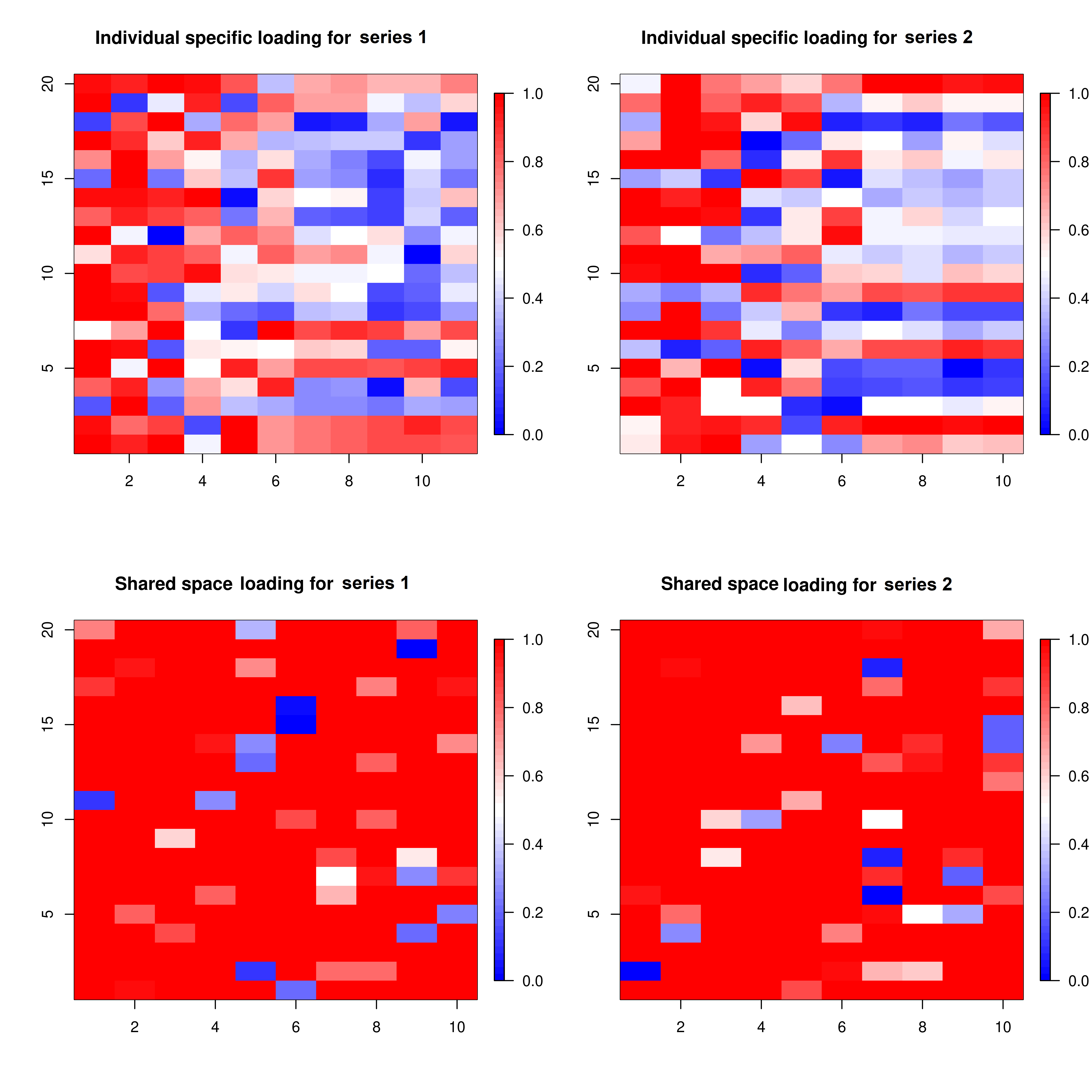}
		\caption{Estimated importance measures $SP$ for loading matrices of shared and individual spaces of Series 1 and 2 in Simulation Case 1. Each column represent each factor. The columns with higher proportion of red correspond to the factors with higher importance.}
		\label{warpingload1}
	\end{figure}

	First, we evaluate whether our model retrieves the appropriate number of factors. The true dimension of $\Lambda$ is $20\times 10$. Figure~\ref{warpingload1} suggests that TACIFA retrieves 10 important shared space factors, as expected. The maximum three principal angles between the true $\Gamma_1$ and $\Lambda$ are $0.39\pi, 0.42\pi$ and $0.47\pi$ %are (1.00, 0.99, 0.97, 0.89, 0.84, 0.75, 0.66, 0.38, 0.26, 0.09) 
	and those between the true $\Gamma_2$ and $\Lambda$ are $0.35\pi, 0.40\pi$ and $0.47\pi$. %(1.00, 1.00, 0.97, 0.92, 0.72, 0.64, 0.55, 0.44, 0.34, 0.08). 
	%The principal angle varies within $[0,\pi/2]$, where $0$ signifies affine subspaces and $\pi/2$ signifies orthogonality. 
	The overall mean for each series belongs to the combined space spanned by the columns of the shared and individual-specific loading matrices. The part not explained by the shared space belongs to the individual-specific spaces for each individual. The true individual-specific loading matrix has only 2 principal vectors that are close to being orthogonal with respect to the principal vectors of the true shared loading matrix $\Lambda_0$. 
	We also have approximately 2 important factors and one moderately important factor in the individual specific spaces in Figure~\ref{warpingload1}. 
	
\begin{figure}[htbp]
		\centering
		\includegraphics[width = 0.35\textwidth]{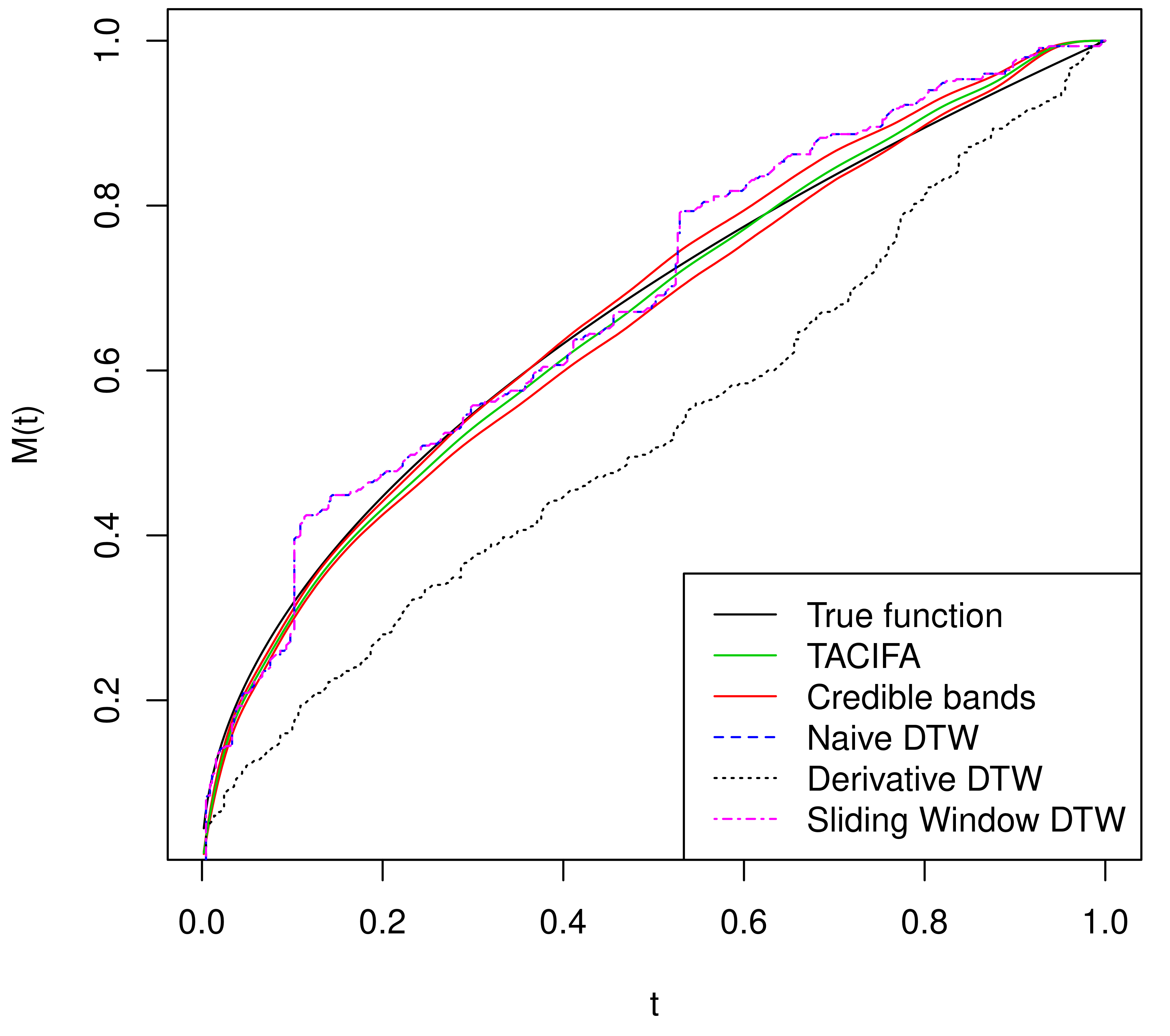}
		\caption{Estimated warping function for simulated data in Simulation Case 1. The black curve is the true function, the green curve is the estimated function, 95\% credible bands are shown in red. Naive DTW and Sliding window DTW curves are indistinguishable.  Of all the methods tested, the TACIFA estimated warping function is closest to the true warping function.}
		\label{warping1}
	\end{figure}

Next, we evaluate the accuracy of our estimated warping function and accompanying uncertainty quantification. The estimated warping function in Figure~\ref{warping1} is for the training set. The estimate by TACIFA is clearly the best among all methods tested. In Table~\ref{sim1}, we compare the prediction MSE results of our method with two-stage methods, and show that TACIFA has the best performance. 
	
	%We also compare the out of sample prediction error of TACIFA and two-stage approaches for estimating shared and individual-specific dynamic factors.  On the test set, we calculate the prediction TACIFA MSEs, which are 1.00 and 1.01 (relative to the true variance of 1) with 95\% and 94\% frequentist coverage within 95\% posterior predictive credible bands for X and Y, respectively. 

		\begin{table}[htbp]
	    \centering
	    \caption{Prediction MSEs of the first and second time series in Simulation 1 using two-stage methods. The top row indicates the R package used to impute, and the first column indicates the warping method. The two-stage prediction MSEs are all greater than the TACIFA prediction MSEs (1.01, 1.02).}
	    \begin{tabular}{|l|l|l|l|}
	        \hline 
	         &{\tt missForest}&{\tt MICE}&{\tt mtsdi}\\
	        \hline
	         Naive DTW&(6.12, 9.66) &(8.65,9.70)  &(1.03,1.03)\\ 
	         Derivative DTW&(6.37, 9.49) &(8.06,9.80)  &(1.03,1.03)\\ 
	         Sliding DTW&(7.15, 10.55) &(9.61,10.39)  &(1.03,1.03)\\ 
	         \hline
	    \end{tabular}

	    \label{sim1}
	\end{table}
	
	Finally we measure the similarity of the simulated data using the measure described in Section~\ref{simmeasure}.   If $\zeta_{1k}(t)=\sin(kt)$ as above, the similarity is 0.95.  To confirm that this measure is sensitive to the similarity between two time series, as intended, we change the first multivariate time series relative to the other multivariate time series by changing the first individual specific latent factors $\zeta_{1k}(t)$ systematically, and recalculating the similarity.  When $\zeta_{1k}(t)=kt$, similarity drops from 0.95 to 0.89.  When $\zeta_{1k}(t)=(kt)^2$, similarity further reduces to 0.79. The warping function estimated for each of these pairs of time series deteriorates as the two multivariate time series become more distinct as expected. Two stage methods do much worse in these cases (Figure 5 of the Supplementary Materials). %This is shown in the Figure 5 of the Supplementary Materials.
	
	\subsection{Simulation case 2}
	In Simulation Case 2, each series reflects a circle changing into an ellipse over time, similar to a mouth gaping and subsequently closing. The area of the shape is kept fixed by modifying the major and minor axis appropriately. The area of an ellipse, with $a$ and $b$ as the lengths of the major and minor axes, is given by $\pi ab$. Thus to have the area remain fixed we need $ab$=constant. We maintain the constant to be 2. With the same true warping function $M_0(t)$ as in the previous simulation, the values for major and minor axes are linked over time across the two individuals. We let $ax(t)=2(t+1)$ where $t$'s are 500 equidistant values between 1/500 and 1 and $bx(t)=2/(t+1)$; here $ax(t)$ and $bx(t)$ are major and minor axes of the ellipse at time $t$ corresponding to $X_t$. At $t=0$, it is a circle. For the second series we then have $ay(t)=2(t^{0.5}+1)$ and $by(t)=2/(t^{0.5}+1)$. We consider the pair of Cartesian coordinates of 12 equidistant points across the perimeter of the ellipse as features (yielding 24 features in total). The features correspond to 12 equidistant angles in $[0, 2\pi)$. Let $\theta_1,\ldots,\theta_{12}$ be those angles. Then $X_{it}=(ax(t)\sin(\theta_i), bx(t)\cos(\theta_i))$ and $Y_{it}=(ay(t)\sin(\theta_i), by(t)\cos(\theta_i))$. 
	
	%As in the earlier simulation case, first we evaluate whether our model retrieves the appropriate number of shared and individual factors. 
	The choices of hyperparameters and the number of MCMC iterations are all the same as in the previous simulation case. We have a pair of 24 dimensional time series. The X or Y coordinate is zero for the following four features $\theta_i=0,\pi$ and $\theta_i=\pi/2, 3\pi/2$. Thus, the warping should not have any effect on these features and should not contribute to the individual-specific space. The remaining 20 features represent 10 features and their mirror images with respect to either the major or minor axis. Thus, we might predict that the shared space should have 10 independent factors, which is consistent with the results displayed in Figure 6 of the Supplementary Materials. As there are 12 features, the individual-specific space should ideally have around two important factors. This is the case for one of the two individual-specific plots in Figure 6 of the Supplementary Materials. For the other individual, there is one more moderately important factor if we set a threshold of 0.9 on the importance measure SP.

	%As in the earlier simulation case, first we evaluate whether our model retrieves the appropriate number of shared and individual factors. The choices of hyperparameters and the number of MCMC iterations are all the same as in the previous simulation case. We have a pair of 24 dimensional time series. For four feature points $\theta_i=0,\pi$ and $\theta_i=\pi/2, 3\pi/2$ either the Y or X coordinate is zero for all the time points. This is why warping will not have any effect on those and contribute to individual specific space. Among the remaining 20 feature points, half of those are the reflections of the remaining half with respect to either major or minor axes. Thus, the shared space should have 10 independent factors which agrees with Figure~\ref{warpingload2}. As there are 12 feature points, the individual specific space should ideally have around two important factors. This is the case for one of the two individual-specific plots in Figure~\ref{warpingload2}. For the other individual, there is one more moderately important factor if we set a threshold of 0.9 on the importance measure $SP$. 
	
\begin{figure}[htbp]
		\centering
		\includegraphics[width = 0.4\textwidth]{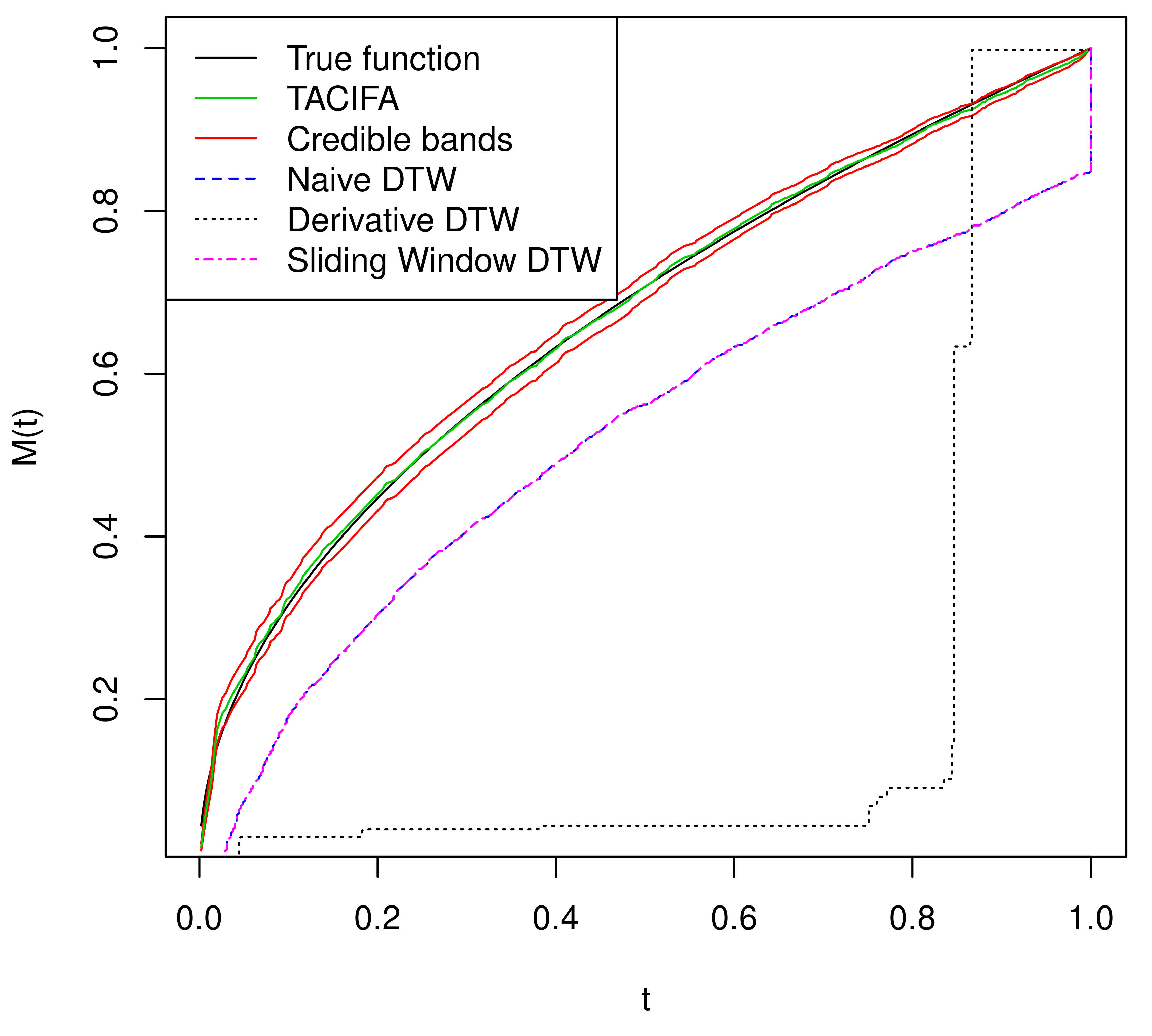}
		\caption{Estimated warping functions for Simulation case 2. The black curve is the true function. The green curve is the TACIFA estimated function, with the TACIFA 95\% credible bands shown in red. Naive DTW and Sliding window DTW curves are indistinguishable. Of all the methods tested, the TACIFA estimated warping function is closest to the true warping function.}
		\label{warping2}
	\end{figure}

\begin{figure}[htbp]
		\centering
		\includegraphics[width = 0.6\textwidth]{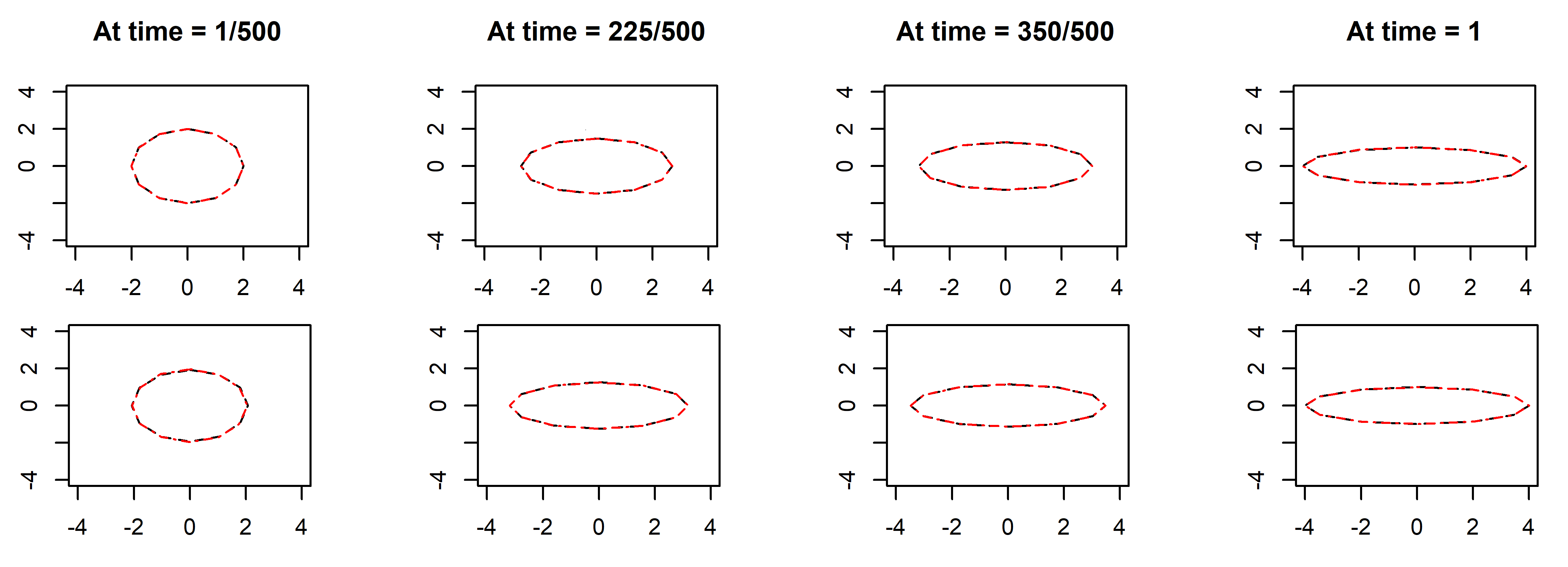}
		\caption{Results for simulation case 2. The black dashed lines represent true curves at four time points and the red dashed lines are the estimated curves. The fit is excellent so that they almost lie on top of each other. At $t=1$, $X$ and $Y$ both have the same shape.}
		\label{progress}
	\end{figure}

	%Next, we compare the accuracy of TACIFA’s estimated warping function and accompanying uncertainty quantification to two-stage approaches.  
	We plot the estimated warping functions in Figure~\ref{warping2}, and plot the estimated shapes in Figure~\ref{progress}.  Figure~\ref{warping2} illustrates that the TACIFA-estimated warping function is once again the most accurate of the tested approaches.  The TACIFA-estimated warping function is almost identical to the true curve, and has tightly concentrated credible bands.  Figure~\ref{progress} confirms that the TACIFA-estimated Cartesian coordinates of the 12 equidistant features are almost perfectly aligned with the true Cartesian coordinates.  Quantifying these accuracies, we calculate the prediction TACIFA MSEs, which are $1.34\times10^{-6}$ and $4.99\times 10^{-6}$ with 95\% and 96\% frequentist coverage within 95\% posterior predictive credible bands for X and Y coordinates, respectively. In Table 2, we compare the results of our method with two-stage methods, and show that TACIFA again has the best performance, this time much more dramatically than in the first simulation.  The method mtsdi gives similar prediction error to our method in the first simulation setup but fails to impute at any of the missing time points for the second simulation. MICE could impute in the first simulation, but only partially for the second simulation. Only missForest could produce results for both of two simulations.  Nonetheless, its prediction MSEs are much higher than those of our method.
	
	\begin{table}[htbp]
	    \centering
	    \caption{Prediction MSEs of the first and second time series in Simulation 2 using two-stage methods. The top row indicates the R package used to impute, and the first column indicates the method used to warp. mtsdi could not impute at any of the testing time points in this simulation. The two-stage prediction MSEs are all greater than the TACIFA prediction MSEs ($1.34\times 10^{-6}$, $4.99\times 10^{-6}$).}
	    \begin{tabular}{|l|l|l|l|}
	        \hline 
	        &{\tt missForest}&{\tt MICE}&{\tt mtsdi}\\
	        \hline
	         Naive DTW&(0.12,0.07) &(0.18,0.09)  &(-,-)\\ 
	         Derivative DTW&(0.12,0.07)  &(0.15,0.07) &(-,-)\\ 
	         Sliding DTW&(0.12,0.07) &(0.14,0.05)  &(-,-)\\ 
	         \hline
	    \end{tabular}

	    \label{sim2}
	\end{table}
	
	\begin{comment}
		\begin{figure}[htbp]
		\centering
		\includegraphics[width = 0.8\textwidth]{sim2load.png}
		\caption{Estimated importance measures SP for loading matrices of shared and individual spaces of Series 1 and 2 in Simulation Case 2. Each column represents a factor. The columns with higher proportion of red correspond to the factors with higher importance.}
		\label{warpingload2}
	\end{figure}
	\end{comment}
	
	%For both of the two simulation cases, the warping function estimation is excellent with the credible bands tightly concentrated around the true curve and the estimated curve due to TACIFA is more accurate than other estimated curves. We can see that in Figure~\ref{progress} the fitted curves align with the true curves almost perfectly even when the data are not generated from a factor model setup. The method {\tt mtsdi} gives similar prediction error with respect to our method in the first simulation setup but fails to impute at any of the missing time points for the second simulation. {\tt MICE} could impute in the first simulation and partially for the second simulation. Only {\tt missForest} could produce results for both of two simulations but the prediction MSE is much higher than those of our method. 
	 
	\section{Human Mimicry Application}
	\label{real}
	We apply TACIFA to data from social interaction experiments. In these experiments, videos are captured while two people interact over Skype. OpenFace software \citep{baltrusaitis2018openface} is used to extract regression scores for the X and Y coordinates of facial features around the mouth, as well as the pitch, yaw, and roll of head positions, in each frame of the video. Here, we apply our method to an experiment where one individual is instructed to imitate the other’s head movement throughout the interaction. We also apply our method to two related experiments with results in Section 3 of Supplementary Materials.
	
	%We consider the time series of $X$ and $Y$ coordinates in the sequence of image frames of 20 facial features around the chin as the dataset for the experiment (A) as the experiment is imitation of head movements. We have all the features, directly related to the experiment. The other two experiments are on smile imitation. Thus we consider most of the lip related features. In total, we consider here six regression scores around the lip and three more predictors on the head position as the data. The predictors on head position are added in the dataset to have some features having no direct connection with the experiment. Smiling does not have any direct connection with the head movement, but with lip features. The six regression scores are provided with openface software \citep{baltrusaitis2018openface} which we also use to collect facial feature data.
	
	The duration of the experiment is rescaled into [0, 1]. The choices of hyperparameters for estimation are kept the same as in the two simulation setups above except for the number of B-splines. We collect 5000 MCMC samples after 5000 burn-in samples. We truncate the columns of the loading matrices that have mean absolute contribution less than $0.0001$. We plot the estimated warping function along with credible bands and the values of $SP(\Psi\Gamma_1), SP(\Psi\Gamma_2), SP(\Lambda\Xi_1),$ and $SP(\Lambda\Xi_2)$ as in the simulation analyses.
	
	%\subsection{Experiment (A)}
	
\begin{figure}[htbp]
		\centering
		\includegraphics[width = 0.8\textwidth]{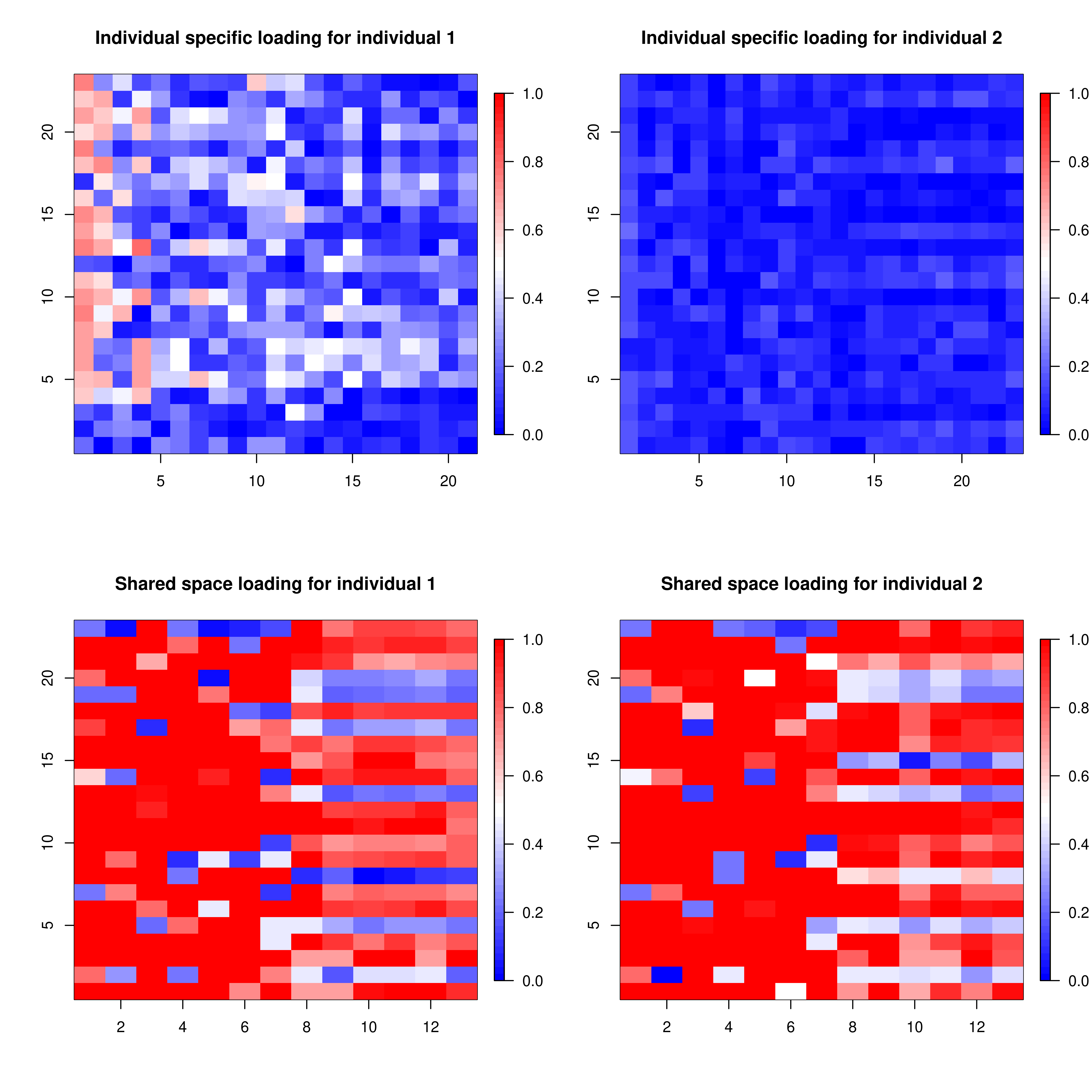}
		\caption{Plot of the summary measure as evidence of importance of the entries of loading matrices in human mimicry dataset (A). Each column represents one factor. The columns with higher proportion of red correspond to the factors with higher importance.}
		\label{loadnochange}
	\end{figure}

	%In this experiment, one individual is instructed to imitate the other’s head movement throughout the interaction. 
	We apply TACIFA to the time courses of 20 facial features from around the mouth and chin along with three predictors of head position. We begin by evaluating the loading matrices of the shared and individual factors. There should be a large shared space in this experiment, as we know one person was imitating the head movements of the other, and all of the features examined were related to the head. We plot $SP(\Psi\Gamma_1), SP(\Psi\Gamma_2), SP(\Lambda),$ and $SP(\Lambda\Xi_2)$ in Figure~\ref{loadnochange}. Half of the 20 facial features examined in this experiment were roughly the mirror image of the others, due to facial symmetry. As a consequence, we might predict that the shared space should not have more than 13 factors. Consistent with this hypothesis, there are 13 important shared features in Figure 7. In addition, all of the features examined in this experiment are related to head movement, so we might predict very little individual variation in the time courses. This prediction is consistent with the low importance of all the individual-specific factors shown in Figure~\ref{loadnochange}.
	
\begin{figure}[htbp]
		\centering
		\includegraphics[width = 0.4\textwidth]{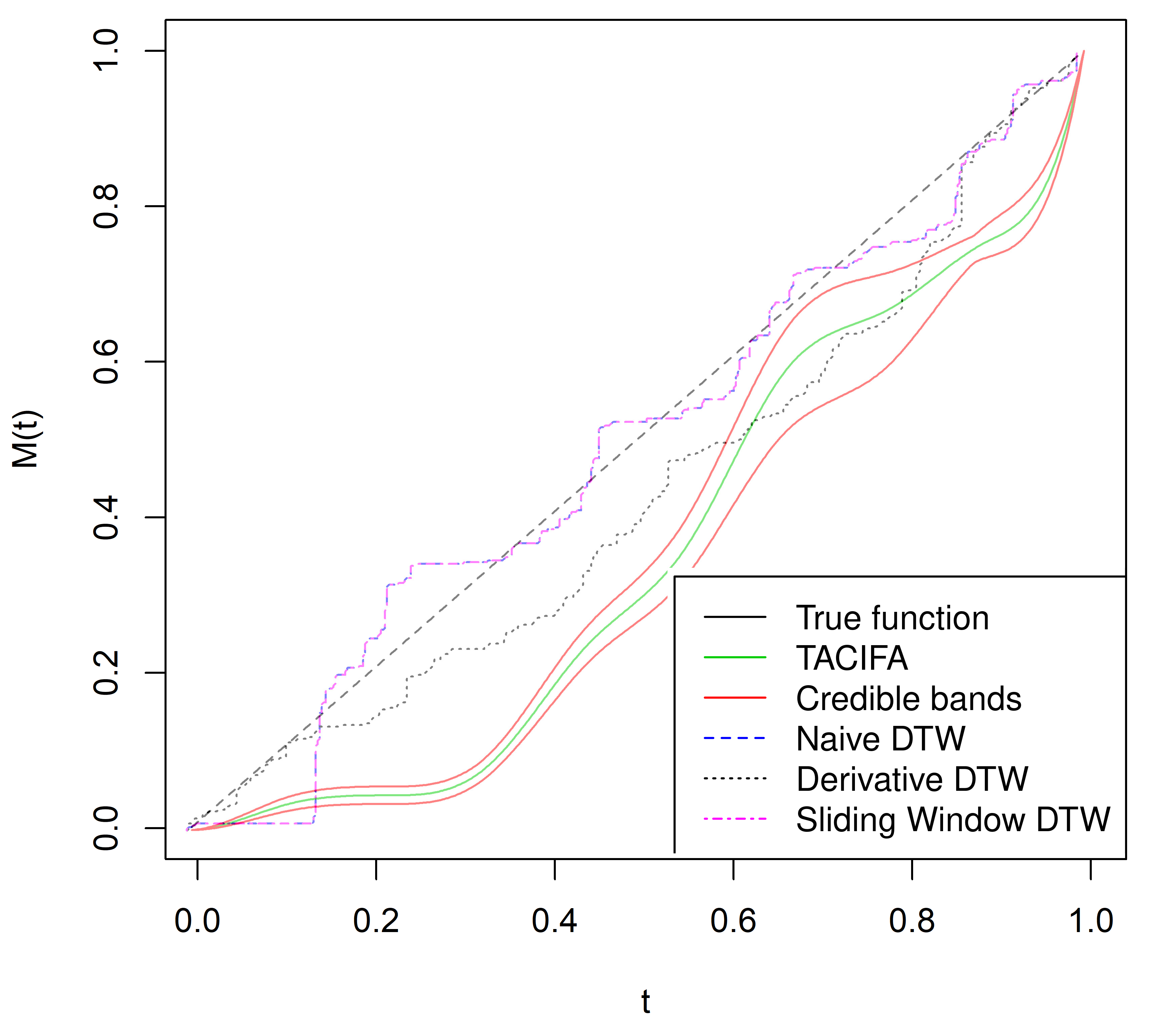}
		\caption{Estimated warping function in human mimicry dataset (A). The green curve is the estimated function along with the 95\% pointwise credible bands in red. The estimated curve is always below the dashed line, indicating the second person is mimicked throughout the experiment}
		\label{real1}
	\end{figure}

	Next, we examine the TACIFA estimated warping function and accompanying uncertainty quantification. Figure~\ref{real1} shows that the estimated warping function is below the $M(t) = t$ line throughout the experiment. This indicates that the TACIFA approach correctly estimated that one individual was following the other individual in time through the experiment. Derivative DTW was the only other method that achieved that.
	
	Next, we compare the TACIFA out of sample prediction MSEs to those of two-stage approaches, and compute the similarity. The TACIFA MSEs are 4.25 and 2.21, with 95\% and 98\% frequentist coverage within 95\% posterior predictive credible bands, relative to the estimated variances 4.34 and 2.61 for the first and second individuals, respectively. These MSEs are lower than those of the two stage approaches, which are around 9. A detailed table is in the Supplementary Materials.
	
	Finally, we assess the similarity of the two time series and test whether greater numbers of features influence the similarity measure. Let $X_m$ and $Y_m$ denote the paired time series with $m$ set of features (maximum of 10) around the chin along with the three predictors on head position. We have a total of 10 possible features in this analysis. We get Syn($X_3, Y_3$)=0.80, Syn($X_6, Y_6$)=0.85 and Syn($X_{10}, Y_{10}$)=0.85. These high values are reasonable. Since all the features examined will be influenced by head movement and head movements were intentionally coordinated. The results also indicate that similarity values increase as the number of relevant features increases.

	\begin{comment}
	\subsection{Similarity measure on shuffled datasets}
	
	We shuffle the time series of one of the two individuals. Then we apply our method and calculate the similarity measure. In this case, we consider all the features for all the three datasets. The similarity measure for all the three real datasets are close to 1. However, for the shuffled dataset (A) it becomes 0.60 and for the other two it is around 0.67. Due to poor estimation of the warping function for the shuffled dataset, the similarity measure is reduced considerably with respect to the corresponding values from the true datasets. 
	\end{comment}

	\section{Discussion}
	\label{discuss}

	There are many possibilities of future research building on TACIFA. It is natural to generalize to $D$ many matrices which would require $D$ different individual-specific loadings $\Gamma_{1},\ldots,\Gamma_{D}$ along with $D-1$ different warping functions. In addition, in settings such as our motivating social synchrony application, there may be data available from $n$ pairs of interacting individuals. In such a case, it is natural to develop hierarchical extension of the proposed approach that can borrow information across individuals and make inferences about population parameters. Another direction is to build static Bayesian models to estimate the joint and individual structures under the orthogonality assumption by dropping the warping function from our proposed model. 
	
	%Although the preliminary results are good, some further theoretical studies on computational efficiency in capturing joint and individual structures will improve the applicability of our method greatly. We have shown good empirical results of our similarity measure and importance measure. However, there is a scope for improvement of these measures using more structures. For example the similarity can be redefined using more information from the model. %In this paper, we proposed a post-process scheme of the MCMC samples of the loading matrix for sake of interpretability. This can further be improved by considering additional sampling variability.
	
%\begin{comment}
\section*{Acknowledgements}
This research was partially supported by
grant R01-ES027498-01A1 from the National Institute of Environmental Health Sciences (NIEHS) of the National
Institutes of Health (NIH).
%\end{comment}

	\bibliographystyle{bibstyle}
	\bibliography{main}
	
	\newpage
	
	 {\LARGE\bf Supplementary Materials for Bayesian time-aligned factor analysis of paired multivariate time series}
	
		\section{Proofs of the theorems}
	
	In this section, the detailed proofs of the theorems are presented.
	
	\subsection{Proof of Theorem 1}
	
	We can write $\|\Lambda_1^{(1)}-\Lambda_1^{(2)}R_1\|_F^2=trace(\Lambda_1^{(1)}-\Lambda_1^{(2)}R_1)^{T}(\Lambda_1^{(1)}-\Lambda_1^{(2)}R_1)=trace\big[(\Lambda_1^{(1)})^{T}\Lambda_1^{(1)})+\Lambda_1^{(2)})^{T}\Lambda_1^{(2)})-2(\Lambda_1^{(1)})^{T}\Lambda_1^{(2)}R_1\big]$, using the properties of trace of a matrix. 
	
	Thus the minimization problem $\|\Lambda_1^{(1)}-\Lambda_1^{(2)}R_1\|_F^2$ is equivalent to the maximization of $trace((\Lambda_1^{(1)})^{T}\Lambda_1^{(2)}R_1)=trace(Q_1DQ_2^{T}R_1)=\sum_{i=1}^rD_{ii}\sum_{j=1}^rx_{ij}y_{ij}$, where $x_{ij}$ and $y_{ij}$ are the $(i,j)$-$th$ entries of $Q_1$ and $(Q_2^{T}R_1)^{T}$. The SVD of $(\Lambda_1^{(1)})^{T}\Lambda_1^{(2)}$ is $Q_1DQ_2^{T}$ with $Q_1$ and $Q_2$ orthonormal matrices. Since $Q_1$ and $(Q_2^{T}R_1)^{T}$ are orthonormal matrices, the above sum is maximized when $x_{ij}=y_{ij}$ for all $(i,j)$ by the Cauchy-Schwarz inequality. Thus $Q_1=(Q_2^{T}R_1)^{T}$ which implies $R_1=Q_2Q_1^{T}$.

	\subsection{Proof of Theorem 2}
	We consider $\Lambda_1=\Lambda\Xi_1$ and $\Lambda_2=\Lambda\Xi_2$. Due to orthogonality between individual-specific and shared space loading matrices, the decomposition into shared space and individual-specific means is identifiable. Let us consider $\Lambda_{11}\eta_1(t)=\Lambda_{12}\eta_2(t)$ and $\Lambda_{21}\eta_1(M_1(t))=\Lambda_{22}\eta_2(M_2(t))$. We consider $M_1$ and $M_2$ to be strictly increasing functions. We assume $M_1$ and $M_2$ are different. By simple arguments we show that this is a contradiction. 
	
	For two full-rank matrices $P$ and $Q$, we have $\Lambda_{11}=\Lambda_{12}P$, $\eta_1(t)=P^{-1}\eta_2(t)$ and $\Lambda_{21}=\Lambda_{22}Q$, $\eta_1(M_1(t))=Q^{-1}\eta_2(M_2(t))$. This implies
	\begin{align}
		Q^{-1}\eta_2(M_2(t)) &= P^{-1}\eta_2(M_1(t)),\nonumber\\
		\eta_2(M_2(t)) &= QP^{-1}\eta_2(M_1(t)).\label{identi}
	\end{align}
	The rank is $r$. The two functions $M_1$ and $M_2$ are increasing, thus they are invertible. This gives us $\eta_2(t) = QP^{-1}\eta_2(M_1\circ M_2^{-1}(t))$. Let $f(t)=M_1\circ M_2^{-1}(t)$ and $f^k$ be $k$ many self convolutions of $f$. Then we have $$\eta_{2}(t)=(QP^{-1})^k\eta_{2}(f^k(t)),$$ for all positive integers $k$.  Such a result also holds for the inverse of $f$, $f^{-1}()$. By construction, $f$ and $f^{-1}$ are monotone. If $f(t)\leq t$, $f^k(t)$ is a decreasing sequence in $k$ for the given $t$. Then $f^{-1}(t)>t$ and the set of its self convolutions is an increasing sequence in $k$. 
	Then we have using product rule limit,
	$$
	\eta_{2}(t)=\lim_{k\rightarrow\infty}(QP^{-1})^k\lim_{k\rightarrow\infty}\eta_{2}(f^k(t))=B\eta_{2}(0),
	$$
	where $\lim_{k\rightarrow\infty}(QP^{-1})^k = B$. Due to boundedness of $\eta_2$, product rule is possible as both of the two limits exist. Similarly we have $\eta_{2}(t) = B^{-1}\eta_{2}(T)$. If $\eta_2(0)=0$ and $\eta_2(T)=0$, then $\eta_{2}(t) = 0$ for all $t$. This is not possible. Also we have $\eta_2(T)=B^2\eta_2(0)$. Thus we must have $\eta_2(0)\neq 0$ and $\eta_2(T)\neq 0$. We then have $M_1(T)=1=M_2(T)$ and $M_1(0)=0=M_2(0)$. Thus for all $t$ such that $f(t)\leq t$, $\eta_2(t)=B\eta_{2}(0)$. Similarly for $f(t)\geq t$, $\eta_2(t)=B\eta_{2}(T)=B^3\eta_2(0)$. For continuity, we must have $B=I$. This is because we have $B^2\eta_2(0)=\eta_2(0)$ which automatically implies $B^2\eta_2(t)=\eta_2(t)$ for all $t$, implying $B=I$. 
	
	Hence, it implies again that $\eta_{2}(t)$ is constant over $t$. Same result holds for $\eta_1$ as well. This is a contradiction as the condition is that $\eta(t)$ is not constant over $t$. Thus the only possibility we have is $M_1=M_2$.

	\subsection{Proof of Theorem 3}
	
	We apply the posterior concentration result from the Section 8.3 of \cite{Ghosal} for non identically distributed observations. It will require verifying the prior concentration in an $\epsilon_n^2$ neighborhood around the truth, existence of exponentially consistent tests for the truth against an alternative in terms of the metric $d_n$ with at most error probabilities $\exp(-c_1n\epsilon_n^2)$, and a ``sieve" in the parameter space with at least $1-\exp^{-c_2n\epsilon_n^2}$ probability that can be covered by at most $\exp(c_3n\epsilon_n^2)$ balls of radius $\epsilon_n$ for some constants $c_1, c_2$ and $c_3$ such that $c_2>c_1+4$. In our model we have $0\leq t_i\leq 1$ for all $1\leq i\leq n$.
	
	For $q, q^* \in$ the space of probability measure $ \mathcal{P}$, let  
	$$
	K(q^*, q) = \int q^*\log{\frac{q^*}{q}} \qquad  V(q^*, q) = \int q^*\log^2{\frac{q^*}{q}}.
	$$
	
	We consider the paired data $(X_i, Y_i)$ as the $i$-$th$ data point. For $\mu_i^* = (\zeta_1^*(t_i)+\eta^*(t_i),\zeta_2^*(t_i)+\eta^*(M^*(t_1))), \Sigma^*=(\Sigma^*_1,\Sigma^*_2)$ $q^*_i=\text{MVN}(\mu_i^*, \Sigma^*)$ and $q_i=\text{MVN}(\mu_i, \Sigma)$, by simple calculations 
	\begin{align*}
		&K(q_i^*,q_i) =\sum_{j=1}^2\sum_{i=1}^p\log\bigg(\frac{\sigma_{ji}}{\sigma^*_{ji}}\bigg) - \frac{1}{2}\bigg[{p}-\sum_{i=1}^p\frac{(\mu_{ij}^*-\mu_{ij})^2}{\sigma_{ji}^2} - \sum_{i=1}^p\frac{\sigma_{ji}^{*2}}{\sigma_{ji}^2}\bigg],\\
		&V(q_i^*,q_i) = \sum_{j=1}^2\sum_{i=1}^p\frac{{1}}{2}\bigg(\frac{\sigma_{jj}^{*2}}{\sigma_{ji}^2}-1\bigg)^2+\sum_{i=1}^p\frac{(\mu_{ij}^*-\mu_{ij})^2}{\sigma_{ji}^4}\sigma_{ji}^{*2}.
		%&d_{H}^2(p_i, q_i) = 1 - \bigg(\frac{2\sigma_1\sigma_2}{\sigma_1^2 + \sigma_2^2}\bigg)^{{J\times T_i}/2}\exp\bigg(-\frac{(\mu_1-\mu_2)'(\mu_1-\mu_2)}{4(\sigma_1^2 + \sigma_2^2)}\bigg).
	\end{align*}
	
	We denote $\mu=(\mu_1,\ldots,\mu_n)$ as all the means stacked together. Let $\mu_0$ and $\sigma_0$ stand for the truth of $\mu$ and $\sigma$. Let $\mu^*$ be such that $\|\mu_0-\mu^*\|>\epsilon_n$. By Lemma 8.27 of \cite{Ghosal} for bounded $\sigma$ one can construct an exponentially consistent test for $(\mu_0,\sigma_0)$ against $(\mu_1,\sigma_1:\|\mu_1-\mu^*\|<\epsilon_n/2)$. Using negative log affinity measure one can construct a test with unbounded support as in \cite{ning2018bayesian}. To keep the proof simple, we consider the additional boundedness assumption.
	
	\begin{align*}
		&d^2((\zeta_{11},\zeta_{21},\eta_1,M_1), (\zeta_{12},\zeta_{22},\eta_2,M_2))\\ &\quad\lesssim n^{-1}\sum_{i=1}^n\big[\|\zeta_{11}(t_i)-\zeta_{12}(t_i)\|_2^2+\|\zeta_{21}(t_i)-\zeta_{22}(t_i)\|_2^2+\|\eta_1(t_i)-\eta_2(t_i)\|_2^2\\&\qquad+\|\eta_1(M_1(t_i))-\eta_2(M_2(t_i))\|_2^2]
	\end{align*}
	where $\lesssim$ stands for inequality up to a constant multiple. Thus to bound $\epsilon_n$-metric entropies, the logarithm of the number of $\epsilon_n$-balls needed to cover a set, we can consider the functions $\eta$ and $M$ separately.
	
	To proceed with the posterior concentration result we consider a sieve in the parameter space of the form, $\mathcal{G}_n=\{\beta_1,\beta_2,\beta, K, J, \sigma:\|\beta_1\|_{\infty}\leq B_{1n},\|\beta_2\|_{\infty}\leq B_{2n},\|\beta\|_{\infty}\leq B_n, K\leq K_n,J\leq J_n, 1/n\leq\sigma_{ij}\leq e^{c'n\epsilon_n^2}, j=1,\ldots,p, i=1,2\}$. We have for two sets of parameters $\Theta = (\zeta_1,\zeta_2,\eta, M, \sigma)$ and $\Theta^* = (\zeta_1^*,\zeta_2^*,\eta^*, M^*, \sigma^*)$,
	$$
	n^{-1}\|\eta(M(t_i))-\eta^*(M^*(t_i))\|_2^2\lesssim B_n^2\max_l|\gamma_l-\gamma_l^*|^2,
	$$
	and
	$$
	n^{-1}\|\eta(t_i)-\eta^*(t_i)\|_2^2\lesssim \max_l|\beta_l-\beta_l^*|^2.
	$$
	The error in approximating the function $\eta$ can be uniformly bounded in an order of $\bar{K}_n^{-\iota}$ using $K_n$ B-spline basis functions. Similarly for $\zeta_1$ and $\zeta_2$ these are up to order $\bar{K}_{1n}^{-\iota_1}$ and $\bar{K}_{2n}^{-\iota_2}$. Also for $M$ with $\bar{J}_n$, it is $\bar{J}_n^{-\iota'}$. Thus we have $\epsilon_n\geq\max(\bar{K}_{1n}^{-\iota_1},\bar{K}_{2n}^{-\iota_2},\bar{K}_n^{-\iota},\bar{J}_n^{-\iota'})$. For our prior $\Pi(\max_l|\gamma_l-\gamma_l^*|\leq c\epsilon_n)>\epsilon^{\bar{J}_n}$, $\Pi(\max_l|\beta_{1l}-\beta_{1l}^*|\leq c\epsilon_n)>\epsilon^{\bar{K}_{1n}}$,$\Pi(\max_l|\beta_{2l}-\beta_{2l}^*|\leq c\epsilon_n)>\epsilon^{\bar{K}_{2n}}$, and $\Pi(\max_l|\beta_l-\beta_l^*|\leq c\epsilon_n)>\epsilon^{\bar{K}_n}$. The contraction rate $\epsilon_n$ must be worse than parametric rate $n^{-1}$. Thus we have $\epsilon^{\bar{K}_{1n}+\bar{K}_{2n}+\bar{K}_n+\bar{J}_n}>\exp(-c'(\bar{K}_{1n}+\bar{K}_{2n}+\bar{K}_n+\bar{J}_n)\log n)$ for some $c'>0$. For a pre-rate $\bar{\epsilon}_n$ we have 
	\begin{equation}
		\label{rate concentration}
		(\bar{K}_{1n}^{-\iota_1}+\bar{K}_{2n}^{-\iota_2}+\bar{K}_n^{-\iota}+\bar{J}_n^{-\iota'})\lesssim \bar\epsilon_n, \quad (\bar{K}_{1n}+\bar{K}_{2n}+\bar{K}_n+\bar{J}_n)\log n\lesssim n\bar\epsilon_n^2.
	\end{equation}
	The actual contraction might be little higher than this pre-rate. It depends on the prior.  
	The bound for $\epsilon_n$-entropy of the sieve will be a constant multiple of $({K}_{1n}+{K}_{2n}+{K}_n+{J}_n)\log n + n\epsilon_n^2$. Taking $B_{1n},B_{2n},B_n$ as a polynomial of $n$, to satisfy the conditions on the sieve according to the general theory, we need
	
	\begin{align}
		\label{rate sieve}
		{K}_{1n}(\log n)^{b_{13}}+{K}_{2n}(\log n)^{b_{23}}+K_n(\log n)^{b_3} + J_n(\log n)^{b_3'}&\gtrsim n\bar\epsilon_n^2,\nonumber\\ \quad ({K}_{1n}+{K}_{2n}+K_n+J_n) e^{-b n^2}&\lesssim \exp[ -n\bar\epsilon_n^2].
	\end{align}
	
	If we consider $\bar{K}_{in}\asymp (n/\log n)^{1/(2\iota_i+1)}$, for $i=1,2$, $\bar{K}_n\asymp (n/\log n)^{1/(2\iota+1)}$, $\bar{J}_n\asymp (n/\log n)^{1/(2\iota'+1)}$ in \eqref{rate concentration}, this leads to $\bar \epsilon_n\asymp (n/\log n)^{-\bar{\iota}/(2\bar{\iota}+1)}$ where $\bar{\iota}=\min\{\iota,\iota_1,\iota_2,\iota'\}$. Now to satisfy~\eqref{rate sieve}, we need $$K_{in}\asymp n^{1/(2\iota_i+1)}(\log n)^{\iota_i/(2\iota_i+1)+(1-b_{i3})/2},$$ for $i=1,2$,
	$$K_n\asymp n^{1/(2\iota+1)}(\log n)^{\iota/(2\iota+1)+(1-b_3)/2},\quad J_n\asymp n^{1/(2\iota'+1)}(\log n)^{\iota'/(2\iota'+1)+1-b_3}.$$ Thus final rate $\epsilon_n$ becomes, 
	$$
	n^{-\bar{\iota}/(2\bar{\iota}+1)} (\log n)^{\bar{\iota}/(2\bar{\iota}+1)+(1-\bar{b_3})/2},
	$$
	where $\bar{\iota}=\min\{\iota,\iota_1,\iota_2,\iota'\}$ and $\bar{b}_3=\min\{b_3,b_3',b_{13},b_{23}\}$.
	
	\subsubsection{Proof of Corollary 1}
	
The empirical $\ell_2$-distance is given by,
	\begin{align*}
		&d_1^2((\Psi_1, \Lambda_1, \Gamma_{11},\Gamma_{12},\Xi_{11},\Xi_{12}, \zeta_{11},\zeta_{21},\eta_1,M_1), (\Psi_2, \Lambda_2, \Gamma_{21},\Gamma_{22},\Xi_{21},\Xi_{22}, \zeta_{12},\zeta_{22},\eta_2,M_2))\\
		&\quad = \frac{1}{n}\sum_{i=1}^n\big[\|\Psi_1\Gamma_{11}\zeta_{11}(t_i)-\Psi_2\Gamma_{12}\zeta_{12}(t_i)\|_2^2+\|\Psi_1\Gamma_{21}\zeta_{21}(t_i)-\Psi_2\Gamma_{22}\zeta_{22}(t_i)\|_2^2\\&\qquad+\|\Lambda_1\Xi_{11}\eta_1(t_i)-\Lambda_2\Xi_{12}\eta_2(t_i)\|_2^2+\|\Lambda_1\Xi_{21}\eta_1(M_1(t_i))-\Lambda_2\Xi_{22}\eta_2(M_2(t_i))\|_2^2\big].
	\end{align*}
	For the original model, test constructions will remain the same. We only need to verify Kullback-Leibler prior positivity. We can show that
	$$
	\|\Psi_1\Gamma_{i1}\zeta_{i1}-\Psi_0\Gamma_{i0}\zeta_{i0}\|_2\leq \|\Psi_1-\Psi_0\|_2+\|\Gamma_{i1}-\Gamma_{i0}(t)\|_2+\|\zeta_{i1}-\zeta_{i0}\|_\infty,
	$$
	for $i=1,2$ and 
	\begin{align*}
	    &\|\Lambda_1\Xi_{11}\eta_1-\Lambda_0\Xi_{10}\eta_0\|_2+\|\Lambda_1\Xi_{21}\eta_1(M_1)-\Lambda_0\Xi_{20}\eta_0(M_0)\|_2\leq\\\quad& \|\Lambda_1-\Lambda_0\|_2+\|\eta_{1}-\eta_{0}\|_\infty+\|M_{1}-M_{0}\|_\infty+\|\Xi_{11}-\Xi_{10}\|_2 + \|\Xi_{21}-\Xi_{20}\|_2.
	\end{align*}
	If the latent factors and loading matrices are close to their true values, then the means of individual-specific and shared space means are close to the corresponding true means. We have already proved that these means converge to their true values in Theorem 3. Thus the Kullback-Leibler divergence converges to zero. This completes the proof of the Corollary.

	\section{Posterior update of $\Gamma_1$, $\Gamma_2$, $\Xi_1,\Xi_2$ and $\beta_1,\beta_2, \beta$}
	
	To update $\Gamma_1$: The posterior mean and variance for $vec(\Gamma_1)$ are given below. Here $vec(\Gamma_1)$ is the vectorized version of the matrix $\Gamma_1$. Let us define a time varying matrix $T(t)$ of dimension $p\times pr_1$ and $i$-$th$ row of $T(t)$ is $T_i(t)=\Psi_i*\zeta_1(t)$, where $*$ denotes convolution of $i$-$th$ row of $\Psi$, and $\zeta_1(t)$. Let $V_{pm}^{\Gamma_1}$ and $M_{pm}^{\Gamma_1}$ denote the posterior variance and mean respectively. Then,
	$$
	V_{pm}^{\Gamma_1}=\Big(\sum_i T_i(t)^{T}\Sigma_1^{-1}T_i(t)+\textrm{diagonal}(vec(PV))\Big)^{-1},
	$$
	where the matrix $PV_{p\times r_1}$ is defined as $PV_{lk}=\Phi_{11,lk}\tau_{11,k}$
	$$
	M_{pm}^{\Gamma_1}=V_{pm}\sum_i T_i(t)^{T}(X_{it}-\Lambda_i\Xi_1\eta(t))
	$$
	
	To update $\Gamma_2$: The posterior mean and variance for $vec(\Gamma_2)$ are given below. Here $vec(\Gamma_2)$ is the vectorized version of the matrix $\Gamma_2$. Define a time varying matrix $T(t)$ of dimension $p\times pr_2$ with the $i$-$th$ row of $T(t)$ being $T_i(t)=\Psi_i*\zeta_2(t)$. Let $V_{pm}^{\Gamma_2}$ and $M_{pm}^{\Gamma_2}$ be the posterior variance and mean respectively. Then,
	$$
	V_{pm}^{\Gamma_2}=\Big(\sum_i T_i(t)^{T}\Sigma_2^{-1}T_i(t)+\textrm{diagonal}(vec(PV))\Big)^{-1},
	$$
	where the matrix $PV_{p\times r_2}$ is defined as $PV_{lk}=\Phi_{12,lk}\tau_{12,k}$
	$$
	M_{pm}^{\Gamma_2}=V_{pm}\sum_i T_i(t)^{T}(Y_{it}-\Lambda_i\Xi_2\eta(M(t)))
	$$
	
	To update $\Xi_1$: The posterior mean and variance for $vec(\Xi_1)$ are given below. Here $vec(\Xi_1)$ is the vectorized version of the matrix $\Xi_1$. Define a time varying matrix $T(t)$ of dimension $p\times pr$ with the $i$-$th$ row of $T(t)$ as $T_i(t)=\Lambda_i*\eta(t)$. Let $V_{pm}^{\Xi_1}$ and $M_{pm}^{\Xi_1}$ be the posterior variance and mean respectively. Then,
	$$
	V_{pm}^{\Xi_1}=\Big(\sum_i T_i(t)^{T}\Sigma_1^{-1}T_i(t)+\textrm{diagonal}(\omega_r)\Big)^{-1},
	$$
	where $\omega_r$ is vector with $\omega$ replicated $r$ times, and
	$$
	M_{pm}^{\Xi_1}=V_{pm}\sum_i T_i(t)^{T}(X_{it}-\Psi_i\Gamma_1\zeta_1(t)).
	$$
	
	To update $\Xi_2$: The posterior mean and variance for $vec(\Xi_2)$ are given below. Here $vec(\Xi_2)$ is the vectorized version of the matrix $\Xi_2$. Let us define a time varying matrix $T(t)$ of dimension $p\times pr$ with the $i$-$th$ row $T_i(t)=\Lambda_i*\eta(M(t))$. Let $V_{pm}^{\Xi_2}$ and $M_{pm}^{\Xi_2}$ be the posterior variance and mean respectively. Then,
	$$
	V_{pm}^{\Xi_2}=\Big(\sum_i T_i(t)^{T}\Sigma_1^{-1}T_i(t)+\textrm{diagonal}(\omega_r)\Big)^{-1},
	$$
	where $\omega_r$ is vector of length $r$ with $\omega$ replicated $r$ times, and
	$$
	M_{pm}^{\Xi_2}=V_{pm}\sum_i T_i(t)^{T}(Y_{it}-\Psi_i\Gamma_2\zeta_2(t))
	$$
	
	For B-spline coefficient matrices $\beta_1,\beta_2$ and $\beta$ of dimensions ${r_1\times K_1}$, ${r_2\times K_2}$ and ${r\times K}$ respectively, we can re-write $\zeta_{1i}(t)=(\chi_t^1)^T\beta_{1i}$,  $\zeta_{2i}(t)=(\chi_t^2)^T\beta_{2i}$ and $\eta_i(t)=(\chi_t)^T\beta_i$ and $\eta(M(t))=(\chi_{M(t)})^T\beta$, where $\chi_t^i$ is the vector of B-spline bases evaluated at time $t$ with $K_i$ many basis functions having equidistant knots. Similarly, $\chi_t$ is the vector of B-spline basis evaluated at time $t$ with $K$ many basis functions having equidistant knots. We can now similarly calculate posterior mean and variances of $vec(\beta_1)$, $vec(\beta)$ and $vec(\beta)$.

	%\subsubsection*{Comparison of the processing}
	
		\section{More Results on Human Mimicry Application}
	
	The detailed table of the comparison between TACIFA and two stage methods from Experiment (A) is given below.
		
			\begin{table}[htbp]
	    \centering
	    \caption{Prediction MSEs of the first and second time series in Experiment (A) using two-stage methods. The top row indicates the R package used to impute, and the first column indicates the method used to warp. mtsdi could not impute at any of the testing time points in this simulation. The two-stage prediction MSEs are all greater than the TACIFA prediction MSEs (4.25 and 2.21).}
	    \begin{tabular}{|l|l|l|l|}
	        \hline 
	        &{\tt missForest}&{\tt MICE}&{\tt mtsdi}\\
	        \hline
	         Naive DTW&(5.21, 7.2) &(17.44, 14.25)  &(5.12, 5.31)\\ 
	         Derivative DTW&(5.21, 7.2)  &(19.15, 15.51) &(5.12, 5.31)\\ 
	         Sliding DTW&(5.21, 7.2) &(17.59, 15.63)  &(5.12, 5.31)\\ 
	         \hline
	    \end{tabular}

	    \label{data1}
	\end{table}
	
		We present results for two additional experiments. In experiment (B), one individual is instructed to imitate the other’s smile for the first part of the experiment, but then the roles are reversed for the second part of the experiment. In experiment (C), the individuals are not doing anything initially, then one individual starts imitating the other’s smile, and then the participants switch roles later in the interaction.

	\subsection{Experiment (B)}
	
	 In this experiment, one individual is instructed to imitate the other’s smile for 55\% of the experiment, and then the roles are reversed for the rest of the experiment. We apply TACIFA and two-stage models to the time courses of six regression scores around the lip and three more predictors on the head position. First we evaluate the loading matrices of the shared and individual factors. Then we assess the ability of TACIFA and two-stage models to identify the known role-reversal at the appropriate time in the experiment, and assess out-of-sample prediction MSEs. Finally, we compute similarity scores using different subsets of predictors.

\begin{figure}[htbp]
		\centering
		\includegraphics[width = 0.8\textwidth]{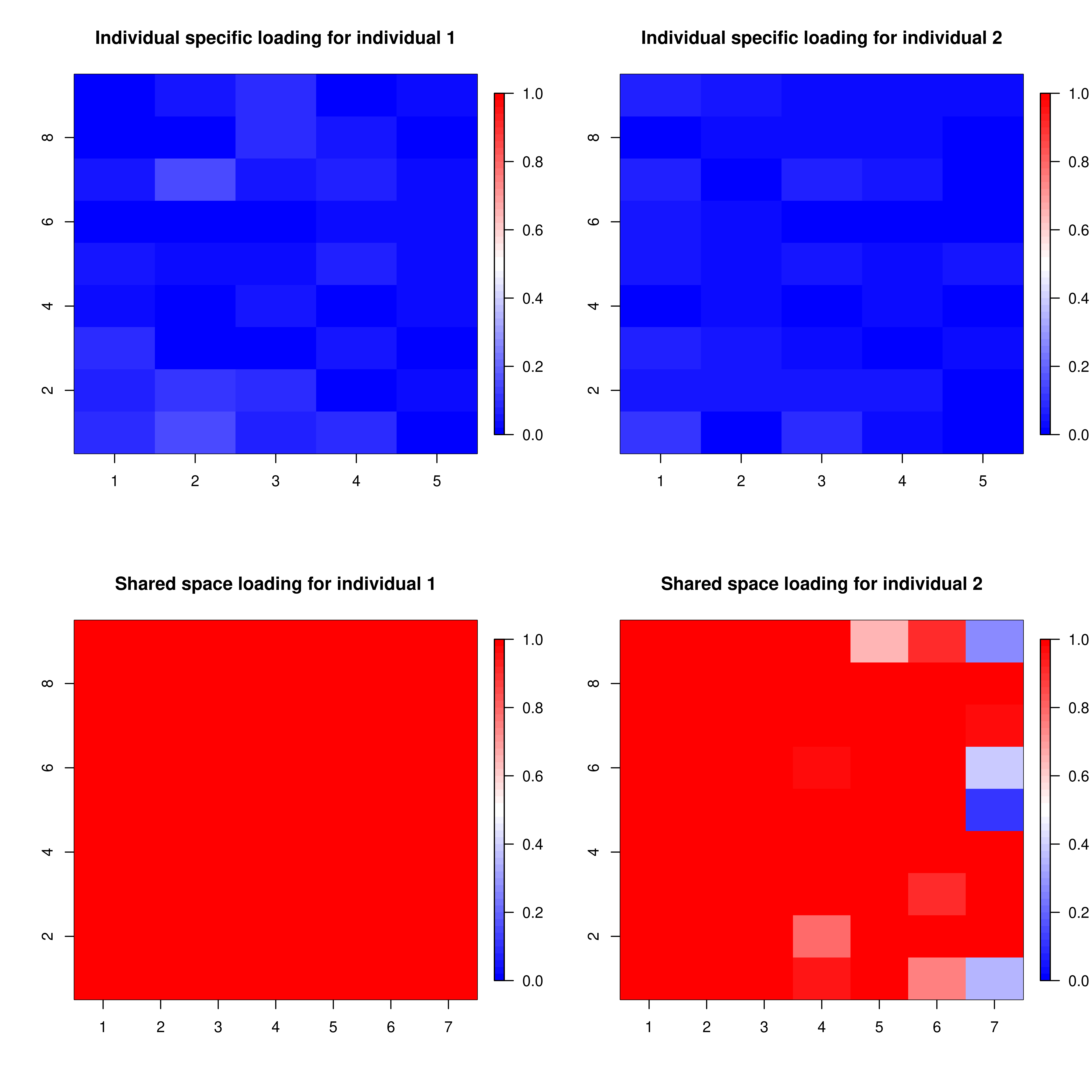}
		\caption{Plot of the summary measure as an evidence of importance of the entries of loading matrices in human mimicry dataset (B). Each column represent each factor. The columns with higher proportion of red correspond to the factors with higher importance.}
		\label{loadchange}
	\end{figure}

The two individuals who participated in this experiment intentionally moved very little outside of the movement associated with smiling. As a consequence, we might predict that there should be very few important individual-specific factors. Figure~\ref{loadchange} confirms this prediction, as none of the individual-specific loading matrices have high SP values. In contrast, 7 components seem to be important in the shared space loading matrices. Since 6 regression based mouth features were included in this model, we might predict that there would be 6 important factors in the shared space. Accordingly, having 7 important factors in the shared space is not unreasonable.

\begin{figure}[htbp]
		\centering
		\includegraphics[width = 0.6\textwidth]{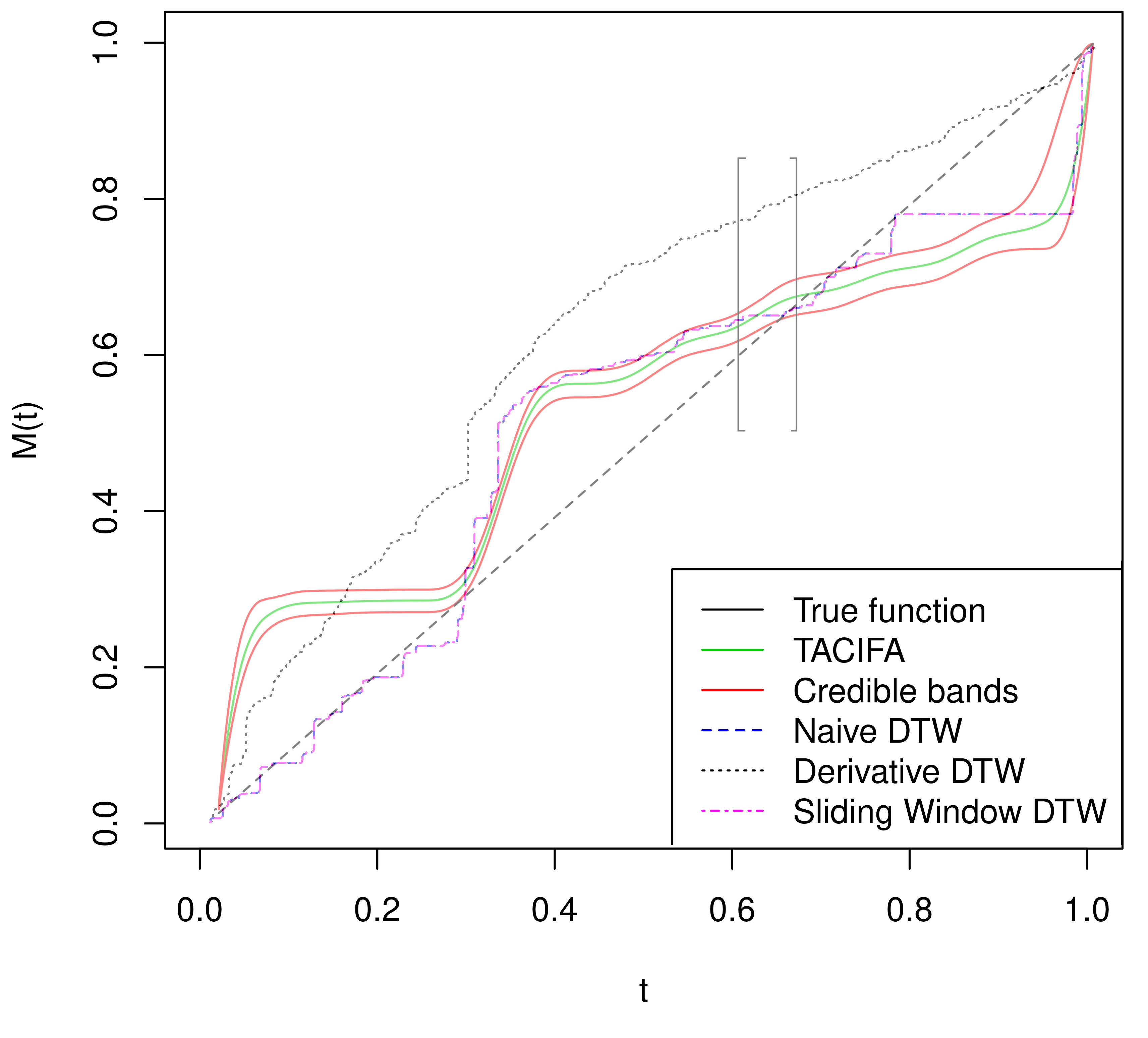}
		\caption{Estimated warping function in human mimicry dataset (B). The green curve is the estimated function along with the 95\% pointwise credible bands in red. The portion of the warping function is marked where the direction of imitation is changed in the data.}
		\label{real2}
	\end{figure}

In addition, TACIFA successfully identifies when the roles of the participants in the experiment switched. In our warping function graphs, the $M(t) = t$ line indicates times when the time points of the shared factor of the two individuals are perfectly aligned. When $M(t) > t$, the first person is leading relative to the second. When $M(t) < t$, the second person is leading relative to the first. Figure~\ref{real2} illustrates the estimated TACIFA warping function with 95\% credible bands, along with the part of the experiment where the direction of mimicry is reversed. For the first part, $M(t) > t$, which correctly implies that the first person is leading, and for the second part of the experiment, $M(t) < t$, which correctly indicates that the second person is leading. Naive DTW and sliding window DTW also show some changes when the roles of the participants in the experiment switched. However, the changes are more prominent for our TACIFA based warping function. Derivative DTW did not detect the direction changes.
	
	The MSEs of out of sample predictions are 0.14 and 0.11 (relative to the estimated variances 0.11 and 0.10), with 88\% and 90\% frequentist coverage within 95\% posterior predictive credible bands, for the first and second individuals, respectively. The prediction MSEs for the two stage methods are all around 0.12 and 0.24 for the two individuals respectively. Detailed results are in Table~\ref{data2}.
	
		\begin{table}[htbp]
	    \centering
	    \caption{Prediction MSEs of the first and second time series in Experiment (B) using two-stage methods. The top row indicates the R package used to impute, and the first column indicates the method used to warp. mtsdi could not impute at any of the testing time points in this simulation. The two-stage prediction MSEs are all greater than the TACIFA prediction MSEs (0.14 and 0.11).}
	    \begin{tabular}{|l|l|l|l|}
	        \hline 
	        &{\tt missForest}&{\tt MICE}&{\tt mtsdi}\\ %(0.04,0.05)
	        \hline
	         Naive DTW&(0.11, 0.18) &(0.15, 0.35)  &(0.10, 0.20)\\ 
	         Derivative DTW&(0.11, 0.19)  &(0.17, 0.33) &(0.10, 0.20)\\ 
	         Sliding DTW&(0.11, 0.18) &(0.17, 0.34)  &(0.10, 0.20)\\ 
	         \hline
	    \end{tabular}

	    \label{data2}
	\end{table}
	
	Finally, we compute the similarity of the time courses extracted from the experiment. Since the participants were intentionally imitating each other’s smiles, the similarity between the time courses should be high when the features in the time courses relate to smiles. The similarity should be lower when the features in the time courses do not relate to smiles. To test the second case, $X_3$ and $Y_3$ denote the paired time series with only head position data, which are not directly related to smiling. To test the first case, $X_6$ and $Y_6$ denote the paired time series with additional 3 smile related features, and $X_9$ and $Y_9$ are the complete data set. We get Syn($X_3, Y_3$)=0.56. However, as would be expected, the similarity increases dramatically when smile-related features are added to the time series, such that Syn($X_6, Y_6$)=0.80 and Syn($X_9, Y_9$)=0.85.

	\subsection{Experiment (C)}
	
	In this experiment, the individuals are instructed not to do anything for the first 25\% of the experiment, then one individual is instructed to imitate the other’s smile, and then the roles are reversed for the rest of the experiment. We again apply TACIFA and two-stage models to the time courses of six regression scores around the lip and three more predictors on the head position as the data. Then we evaluate the same set of metrics as in experiment (B).
	
\begin{figure}[htbp]
		\centering
		\includegraphics[width = 0.8\textwidth]{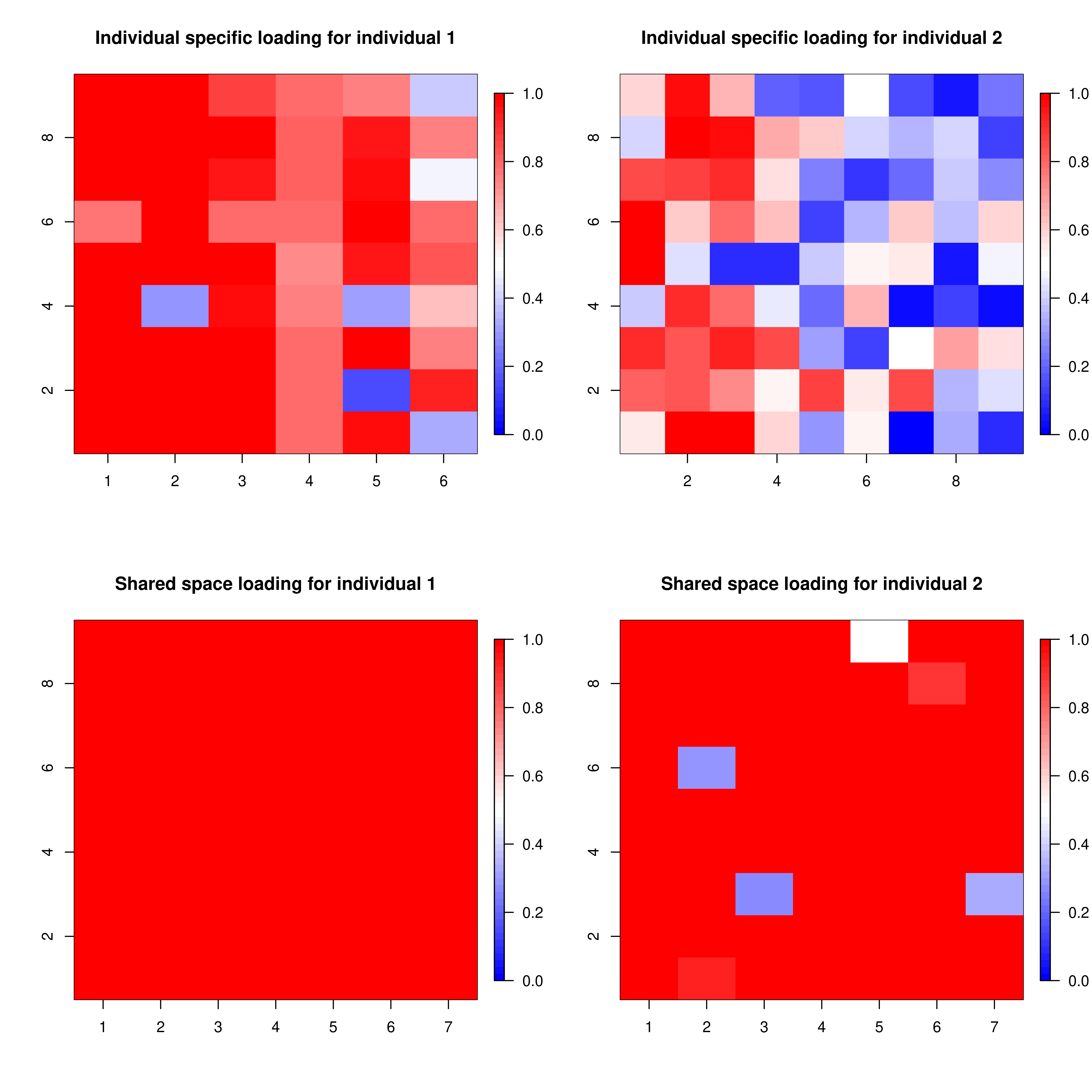}
		\caption{Plot of the summary measure as an evidence of importance of the entries of loading matrices in human mimicry dataset (C). Each column represent each factor. The columns with higher proportion of red correspond to the factors with higher importance.}
		\label{loadnonechange}
	\end{figure}

	In Figure~\ref{loadnonechange}, the first three components for individual 1 and individual 2 of the individual-specific loading matrices seem to be important, and 7 components seem to be important in the shared space loading matrices. Having three individual specific factors is consistent with having three predictors on head position that are unrelated to the experiment of smile imitation. Again, we might predict that there would be 6 important factors in the shared space. Thus, again , having 7 important factors in the shared space is not unreasonable.
	
\begin{figure}[htbp]
		\centering
		\includegraphics[width = 0.6\textwidth]{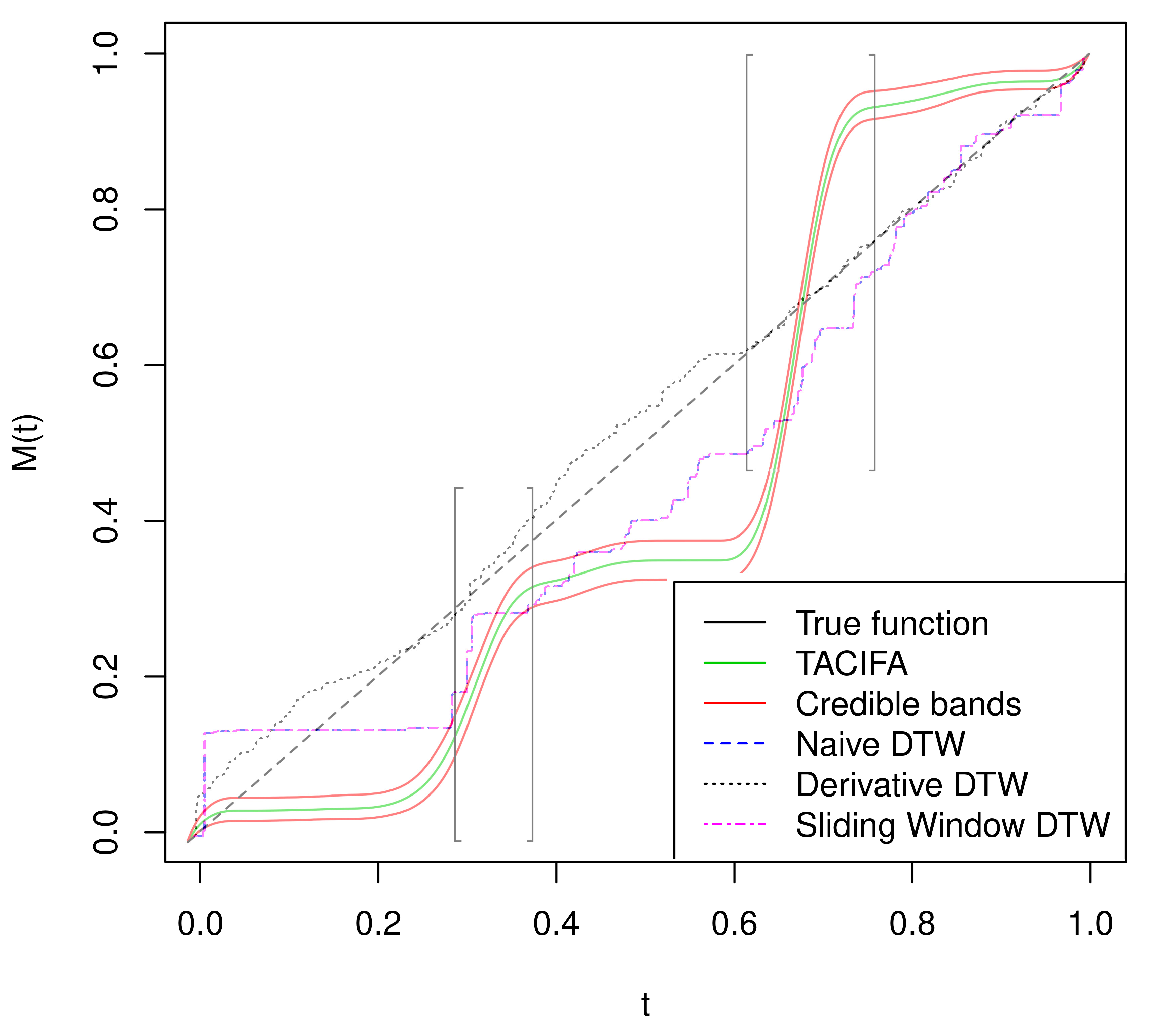}
		\caption{Estimated warping function in human mimicry dataset (C). The green curve is the estimated function along with the 95\% pointwise credible bands in red. The portions of the warping function are marked where in the first marked interval the participants start imitating and then in the next marked interval the direction of imitation is changed in the data.}
		\label{real3}
	\end{figure}

	We also find that TACIFA identifies the change points of the experiment more accurately than other methods in Figure~\ref{real3}. The estimated warping function along with credible bands are shown is Figure~\ref{real3}. The interval where the participants start to imitate is marked in this plot as well as the interval where the direction of mimicry is switched. For the initial few time points, the warping function is flat %, signifying no association between the two time series 
	and at around 75\% time it intersects the dashed line. There are some changes in naive DTW and sliding window DTW curves at these change points on the experiment. However, the changes are more prominent for our TACIFA based warping function in this case also. Once again, derivative DTW did not detect the mimicry direction changes at all.

	The MSEs of out of sample predictions are 0.21 and 0.18 (relative to the estimated variances 0.19 and 0.11) with 88\% and 90\% frequentist coverage within 95\% posterior predictive credible bands for the first and second individuals, respectively. The prediction MSEs for the two stage methods are all around 0.17 and 0.35 for the two individuals respectively. Detailed results are in Table~\ref{data3}.

	\begin{table}[htbp]
	    \centering
	    \caption{Prediction MSEs of the first and second time series in Experiment (C) using two-stage methods. The top row indicates the R package used to impute, and the first column indicates the method used to warp. mtsdi could not impute at any of the testing time points in this simulation. The two-stage prediction MSEs are all greater than the TACIFA prediction MSEs (0.21 and 0.18).}
	    \begin{tabular}{|l|l|l|l|}
	        \hline 
	        &{\tt missForest}&{\tt MICE}&{\tt mtsdi}\\ %(0.02, 0.03)
	        \hline
	         Naive DTW&(0.14, 0.28) &(0.23, 0.36)  &(0.14, 0.30)\\ 
	         Derivative DTW&(0.14, 0.28)  &(0.30, 0.46) &(0.14, 0.30)\\ 
	         Sliding DTW&(0.14, 0.28) &(0.31, 0.45)  &(0.14, 0.30)\\ 
	         \hline
	    \end{tabular}

	    \label{data3}
	\end{table}

	Finally, we computed the similarity between the time series of the two individuals. The definitions of $X_3, Y_3, X_6, Y_6, X_9$ and $Y_9$ are same as in the previous subsection. We begin by comparing the similarity during the part of the experiment where the individuals were not instructed to imitate each other to the part of the experiment where the individuals did imitate each other. Syn($X_{9},Y_{9}$)=0.67 for the non-imitation section and Syn($X_{9},Y_{9}$)=0.84 for the imitation section, so the similarity increased, as would be predicted. Next, we tested whether similarity increased when smile-related features are added to the time series, as in the last experiment. We obtain Syn($X_3,Y_3$)=0.63, Syn($X_6,Y_6$)=0.80 and Syn($X_{9},Y_{9}$)=0.85, suggesting similarity does increase as smile-related features are added.

		\begin{figure}[htbp]
		\centering
		\subfigure[For the case: $\zeta_{1k}(t)=kt$]{\label{fig:a}\includegraphics[width=50mm]{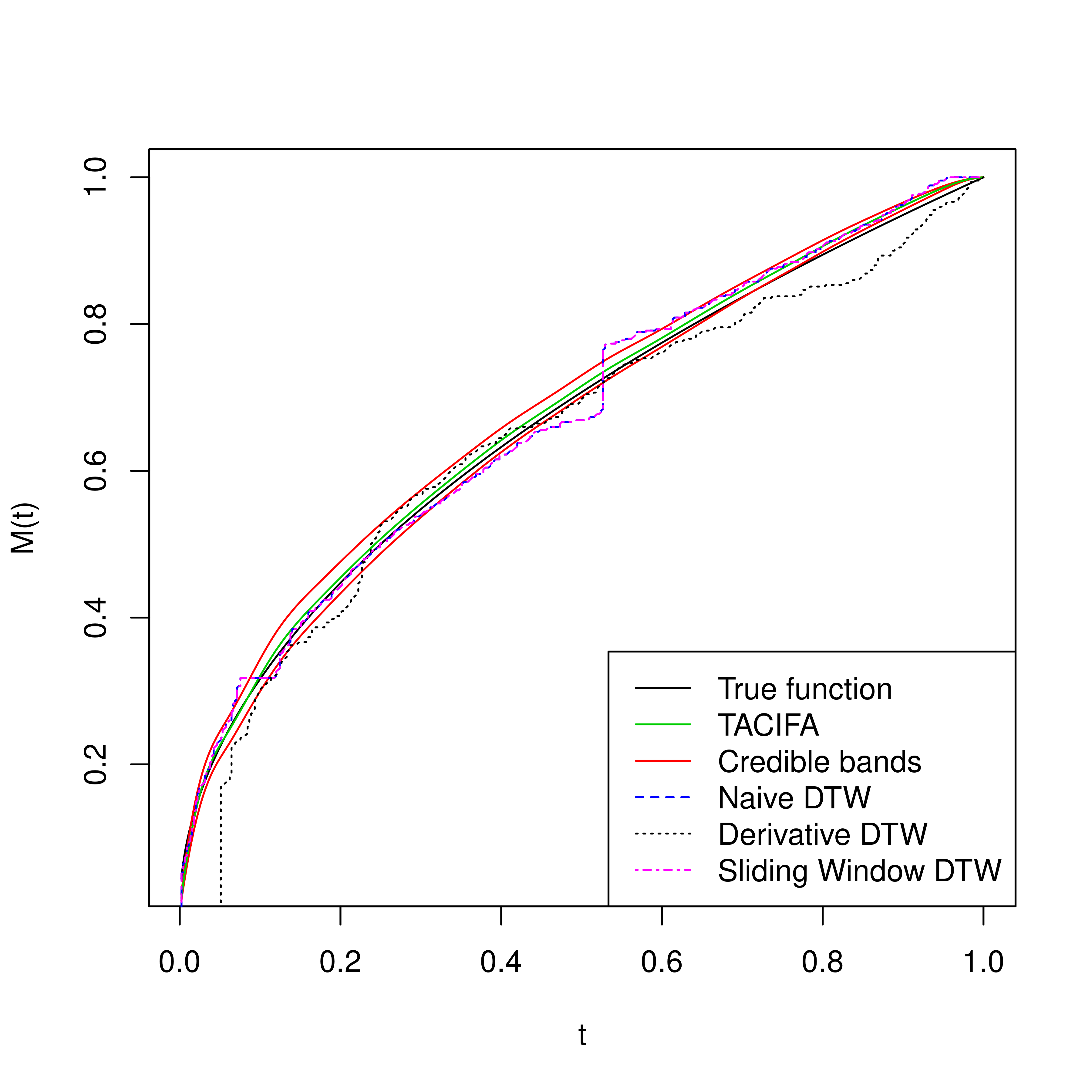}}
		\subfigure[For the case: $\zeta_{1k}(t)=(kt)^2$]{\label{fig:b}\includegraphics[width=50mm]{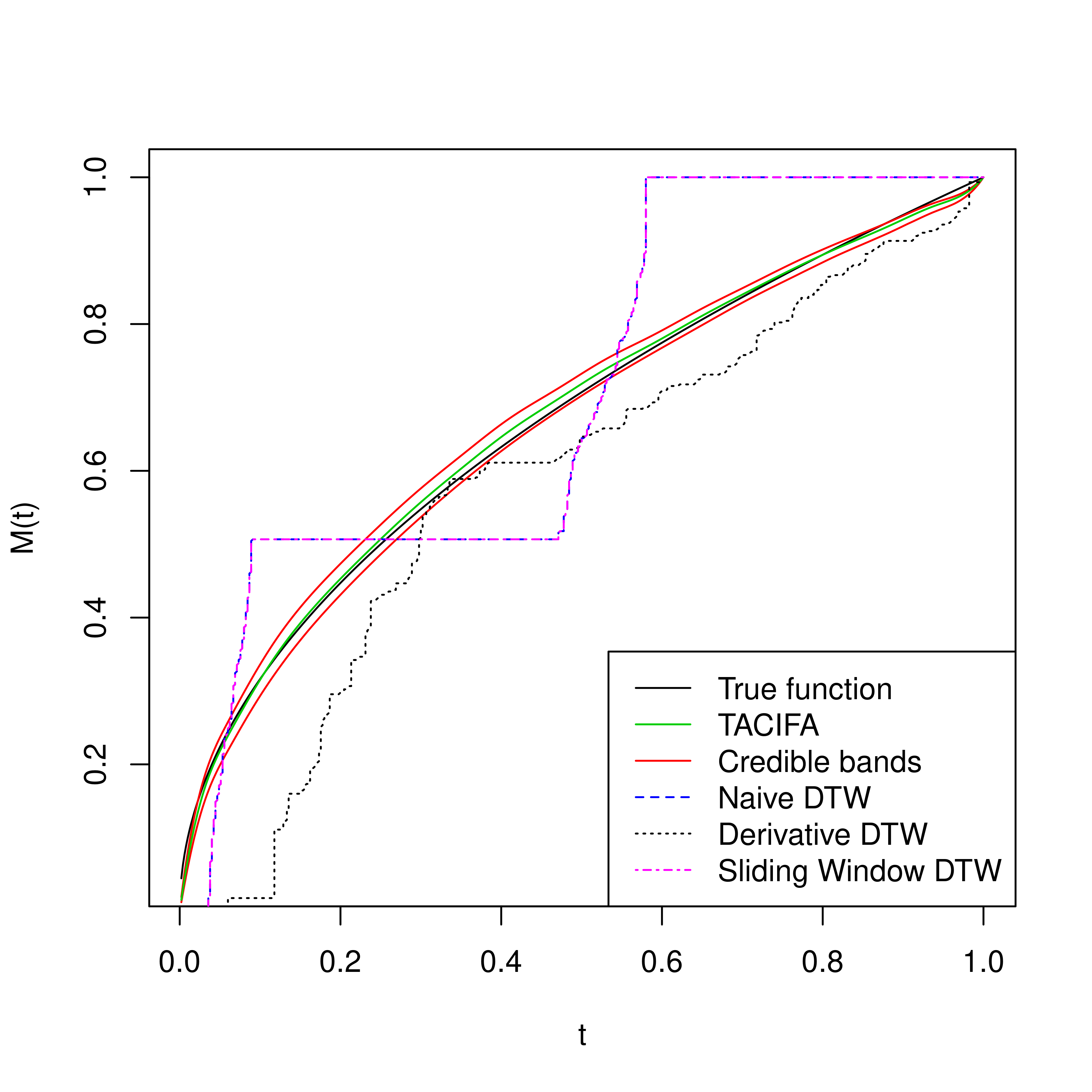}}
		\subfigure[For the case: $\zeta_{1k}(t)=(kt)^3$]{\label{fig:b}\includegraphics[width=50mm]{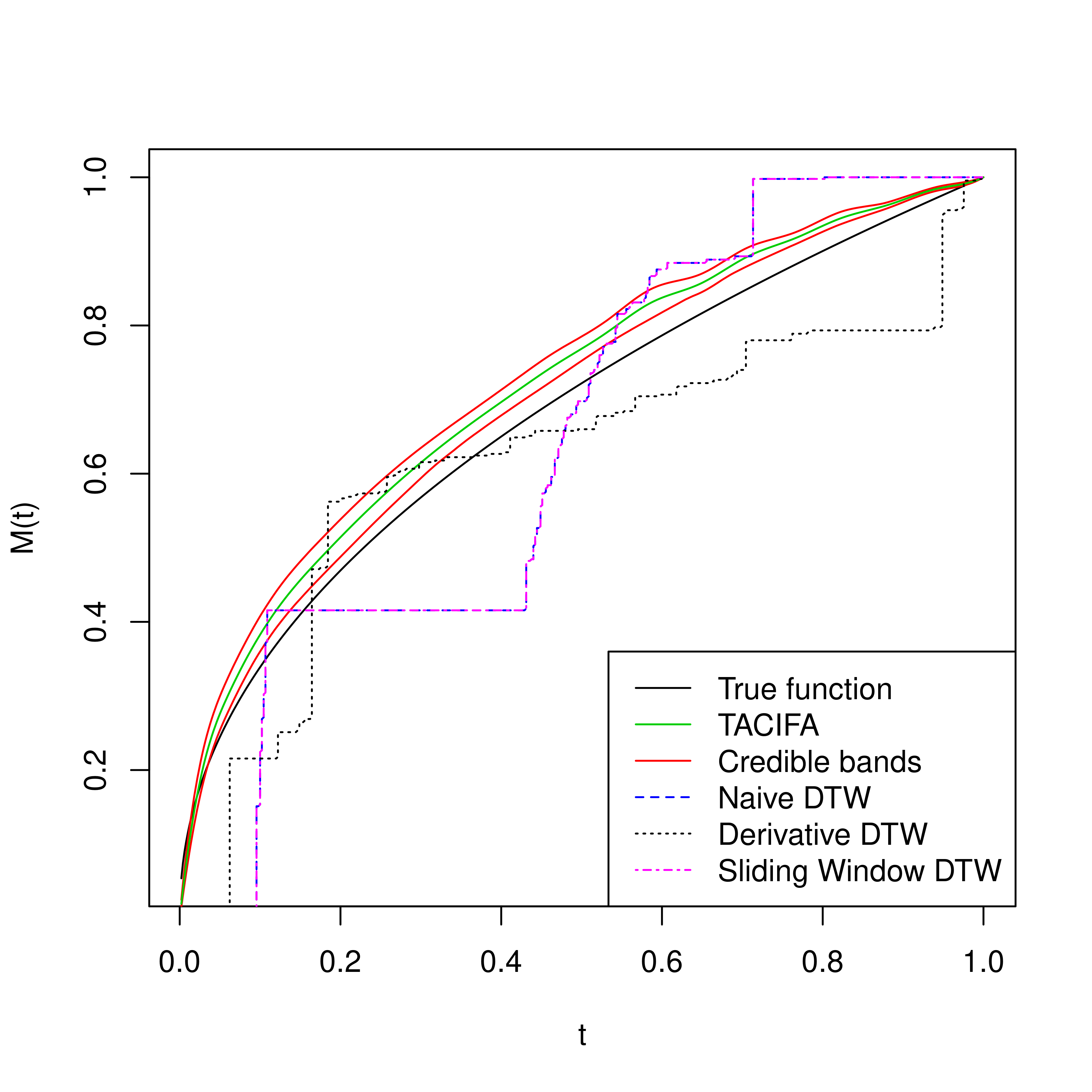}}
		\caption{Estimated warping functions for different choices of $\zeta_{1k}(t)$. In all of these plots, TACIFA estimated warping is the best. However, as the power in $(kt)$ increases, the estimates are getting worse.}
		\label{warpingex}
	\end{figure}
	
		\begin{figure}[htbp]
		\centering
		\includegraphics[width = 0.8\textwidth]{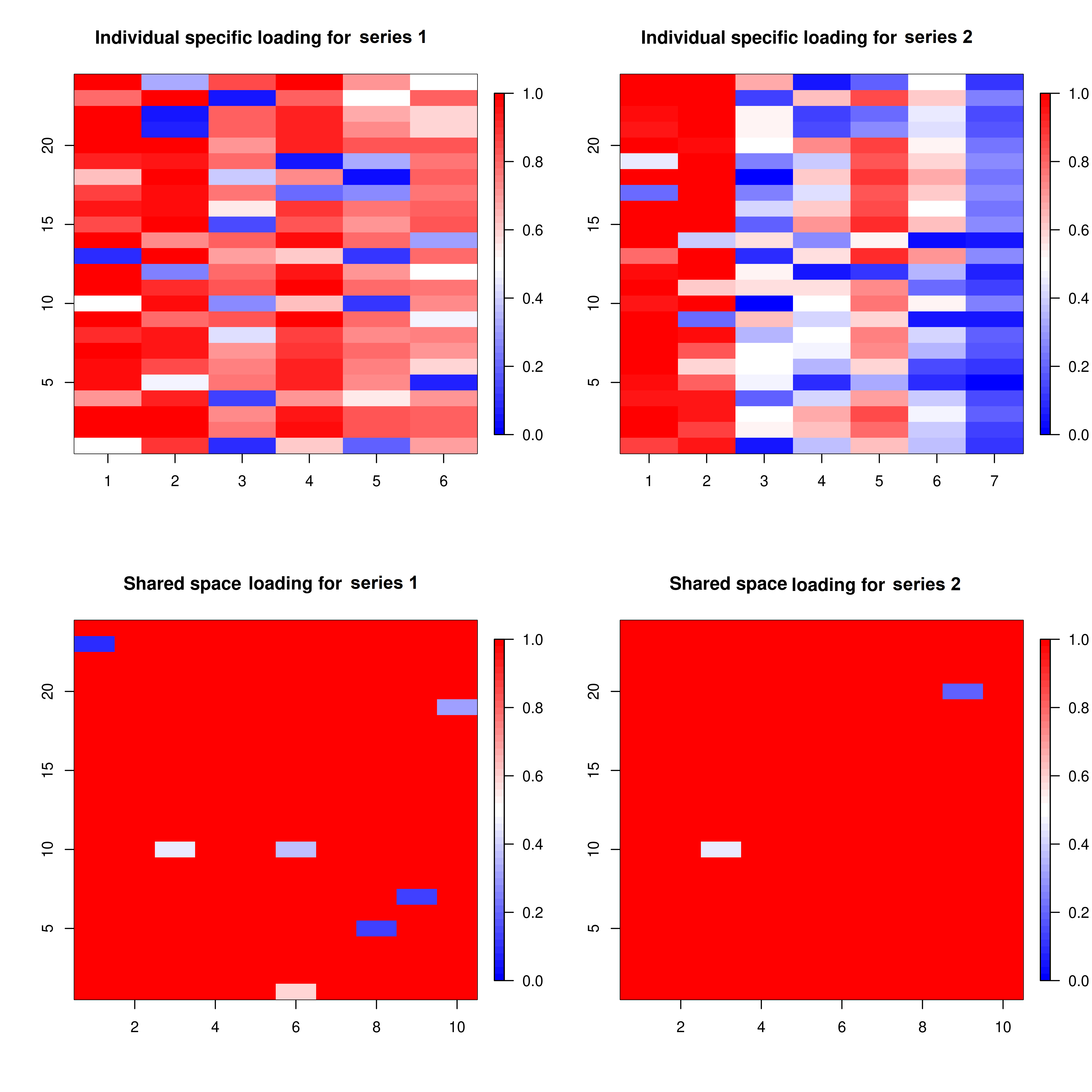}
		\caption{Estimated importance measures SP for loading matrices of shared and individual spaces of Series 1 and 2 in Simulation Case 2. Each column represents a factor. The columns with higher proportion of red correspond to the factors with higher importance.}
		\label{warpingload2}
	\end{figure}

\end{document}